\documentclass[reprint,aps,prapplied, floatfix]{revtex4-2}
\usepackage{xcolor}
\usepackage[normalem]{ulem}
\usepackage{amsmath}
\usepackage{amssymb}
\usepackage{graphicx}
\usepackage{dcolumn}
\usepackage{bm}

\begin{document}

\preprint{APS/123-QED}

\title{Optimizing Pulsed Optically Detected Magnetic Resonance Acquisition: A Digital Signal Processing Perspective}

\author{Anuvab Nandi$^{1}$, Sayan Chakraborty$^{1}$, Yashvardhan Jain$^{1,2}$, Kanav Sharma$^{1,3}$, Samiran Chakraborti$^{1,4}$, Himadri Himani$^{5,6}$, Abir Mondal$^{1,7}$, Sumit Mukherjee$^{1,8}$, Chiranjib Mitra$^{1}$
}

\affiliation{$^{1}$Department of Physical Sciences, Indian Institute of Science Education and Research Kolkata, Mohanpur 741246, West Bengal, India\\
$^{2}$Department of Physics, Boston College, Chestnut Hill, Massachusetts 02467, USA\\
$^{3}$Department of Electromagnetism and Nanoelectronics, Lund University, Box 117, SE-221 00 Lund, Sweden\\
$^{4}$Department of Physics and Astronomy, University of Pittsburgh, Pittsburgh, Pennsylvania 15260, USA\\
$^{5}$Department of Physics, Indian Institute of Space Science and Technology, Thiruvananthapuram 695547, Kerala, India\\
$^{6}$TUM School of Natural Sciences, Technical University of Munich, 85748 Garching, Germany\\
$^{7}$Institute for Quantum Optics, Ulm University, 89081 Ulm, Germany\\
$^{8}$Department of Astrophysics and High Energy Physics, S. N. Bose National Centre for Basic Sciences, Kolkata 700106, India
}

\begin{abstract}

Ensembles of nitrogen vacancy (NV) defect centers in diamond have attracted considerable interest for fundamental studies and applications in quantum sensing and quantum information processing. Pulsed optically detected magnetic resonance (ODMR) and related pulse-sequence protocols are essential for the coherent manipulation and characterization of the NV spins. Here, we introduce a novel framework for analyzing pulsed ODMR data acquisition and subsequent processing from the perspective of digital signal processing (DSP). As a central part of this approach, we model the data processing procedure as an effective digital bandpass filter that suppresses a broad range of noise spectral components, thereby reducing the noise in the processed data. This analysis identifies an intermediate frequency (IF) band whose spectral contribution to this noise is maximum. The proposed DSP-based analysis is experimentally validated using an oscilloscope based acquisition method, and is also applied to optimize key measurement parameters. In particular, the introduction of an analog low pass filter in the signal path is shown to remove aliasing of noise into the IF band, and thus improve signal quality. Our results also establish the oscilloscope as a viable data acquisition device for pulsed ODMR, with practical advantages including on-board waveform averaging, real-time signal visualization, and simplified  system debugging. The proposed framework can be extended to other detection techniques and material platforms for ODMR. 
   
\end{abstract}

\maketitle

\section{\label{sec:intro}Introduction }

Optically active, atom-like defect centers in various solid state materials have emerged as promising candidates for applications in quantum technology \cite{childress2014atom,seo2017designing,awschalom2018quantum,ruf2021quantum,aharonovich2022quantum,castelletto2020silicon,khoury2022bright}. Among these, the negatively charged nitrogen vacancy center in diamond is one of the most extensively studied systems. Its long spin coherence time at room temperature, together with convenient optical initialization and readout, and straightforward microwave control, makes this system an attractive candidate for a wide range of quantum applications \cite{sewani2020coherent}.

Both single isolated defects and ensembles of NV centers are widely used\cite{nizovtsev2005quantum,humphreys2018deterministic,acosta2009diamonds}, depending on the specific application. Ensembles of NV centers have emerged as a promising quantum sensor for detecting magnetic field\cite{Balasubramanian2019}, electric field\cite{Michl2019}, temperature\cite{Choe2018} etc. While a single NV center offers superior spatial resolution, it has a weaker optical signal, lower sensitivity and requires more complex instrumentation\cite{Allert2022}. Sensors based on ensembles of NV centers have found a wide range of applications spanning biological systems, condensed matter physics, electrical and electronic systems, and many more\cite{schirhagl2014nitrogen,ho2022diamond,basu2025diamond}. These sensors, initially demonstrated in bulky benchtop experiments, are increasingly being miniaturized, bringing them closer to practical and commercial deployment\cite{pogorzelski2024compact,wu2025fiber,hiroshige2023compact}. In parallel, NV ensembles have been widely utilized to investigate a range of fundamental phenomena in physics including quantum thermodynamics\cite{bar2019nv,klatzow2019experimental}, open quantum systems \cite{qiu2015quantum,stanwix2010coherence}, and quantum many-body physics\cite{choi2017observation,he2023quasi,wu2025spin}, among others. Beyond fundamental studies, NV ensembles provide a versatile platform for the rapid development and benchmarking of novel pulse sequences, which can subsequently be adapted for single NV center implementations. The aforementioned applications are enabled by optically detected magnetic resonance (ODMR), an experimental technique for probing the electronic spin states of NV centers. In particular, pulsed ODMR provides a versatile platform for coherent control and characterization of NV spins. 

ODMR involves detecting changes in the photoluminescence (PL) from the defect centers induced by the application of resonant microwaves.
The optical signal can be detected using either an analog avalanche photodetector (APD) or a digital single photon avalanche photodiode (SPAD). The analog APD is often the preferred choice, as its dynamic range is high enough to accommodate the optical signal from the NV ensemble. The analog output from the photodetector is digitized with the help of an analog to digital converter (ADC),  within a data acquisition (DAQ) system, a standalone  digitizer, or an oscilloscope \cite{babashah2023optically}. The acquired data are subsequently processed to obtain the final ODMR results. In this work, we develop a novel framework based on digital signal processing (DSP) to systematically analyze the data acquisition and processing sequence, and quantify its effect on the resulting ODMR data. Although the framework is developed and demonstrated using an oscilloscope-based acquisition, it can be readily extended to other acquisition devices, and any other system exhibiting ODMR. Using the results and insights provided by this framework, we modified the experimental setup and optimized the measurement parameters to obtain high-quality ODMR data in short acquisition times.

The use of a digital oscilloscope offers several practical advantages. It enables testing, characterization, debugging and configurable data-acquisition, to be carried out using a single instrument. During the course of setup development, it facilitates verification of the performance of individual components as well as intermediate sub-assemblies. The oscilloscope is particularly useful when designing new protocols, as the pulse sequences can be visually inspected and validated prior to measurements. Moreover, in the event of experimental anomalies, the oscilloscope provides rapid and intuitive means to diagnose and resolve problems. In addition to these diagnostic capabilities, the oscilloscope has features that can be advantageous for acquisition of pulsed ODMR data. In particular, it enables real-time visualization of the signal, as demonstrated in the Supplementary Materials. Its on-board waveform averaging allows for more efficient data acquisition. Although the use of oscilloscope for pulsed ODMR data acquisition has been reported previously\cite{babunts2023magnetic}, there are hardly any descriptions of the corresponding measurement protocol in literature. In this work, we address this gap by applying our signal processing framework to develop a systematic approach towards pulsed ODMR data acquisition using an oscilloscope. 

The paper is organized as follows. Section \ref{sec:sig_pros_not} introduces the signal processing notations used throughout this work. Section \ref{sec:methods} describes the detection part of the experimental setup, followed by the acquisition of the raw data using the oscilloscope, and the subsequent processing steps used to obtain the final ODMR data. Section \ref{sec:sig_noise} examines the signal-bearing components as well as the noise components of the data. We use measures such as standard deviation and Allan deviation to quantify different aspects of the noise in the acquired data. We subsequently examine how the spectral components of the noise affect these measures. In order to quantitatively assess the signal quality of the processed data, we introduce the contrast to noise ratio (CNR). In Section \ref{sec:noise_filt}, we show that the data processing method can be modeled as an effective band pass filter. By suppressing noise over a broad spectral range, the processing protocol reduces the noise in the final data substantially. This analysis allows us to identify an intermediate frequency (IF) band in the noise spectrum which is primarily responsible for the remnant noise in the processed data.

This framework is subsequently used to study the dependence of CNR on various measurement parameters. Section \ref{sec:wav_avg} examines waveform averaging, modeling it as a digital filter to explain the improvement in CNR with number of averages. In Section \ref{sec:readout}, the readout duration - an important experimental parameter - is tuned to achieve the highest CNR. Section \ref{sec:samp_rate} explores how the sampling rate and noise bandwidth collectively influence the CNR. The analysis enabled us to experimentally demonstrate that the introduction of a low pass filter allows more efficient acquisition of the data by suppressing aliasing of noise into the IF band.

\section{Signal Processing Notation}\label{sec:sig_pros_not}

\begin{table*}
\caption{\label{tab:table1}This table summarizes the signal processing notation used throughout this work, alongside the conventional notation and the correspondence between the two. }
\begin{ruledtabular}
\begin{tabular}{ccccc}

 Quantity &Our Notation & Standard Notation & Connection
\\ \hline
 Analog Signal &$x(t)$ & $x(t)$ & NA \\
 Discrete-time Digital Signal  &$x[t_n]$
 &$x[n]$&$t_n=n\cdot T_s$\\
 Fourier transform of Analog Signal&$X(f)$&$X(f)$
 &$NA$ \\
 DTFT of Digital Signal&$\widetilde{X}(f)$&$X(e^{i \omega_d})$ & $\omega_d=2 \pi f/F_s$\\
 DFT of Digital Signal& $\widetilde{X}[f_k]$&$X[k]$& $f_k=k F_s/N$ \\
 Noise in analog Signal & $\delta x(t)$ & $\delta x(t)$ & NA \\
 Noise in Digital Signal & $\delta x[t_n]$&$\delta x[n]$ & $t_n=n\cdot T_s$\\
 PSD of Analog Noise & $S_{\delta x}(f)$ & $S_{\delta x}(f)$ & NA
 \\ DTFT based PSD of Digital Noise & $\widetilde{S}_{\delta x}(f)$ & $S_{\delta x}(e^{i\omega_d})$ & $\widetilde{S}_{\delta x}(f)=S_{\delta x}(e^{i \omega_d}=e^{i
2 \pi f/F_s})/F_s$
 \\DFT based PSD of Digital Noise & $\widetilde{S}_{\delta x}[f_k]$ & $S_{\delta x}[k]$ & $\widetilde{S}_{\delta x}[f_k=k F_s/N]=S_{\delta x}[k]/{F_s}$

\end{tabular}
\end{ruledtabular}
\end{table*}

In this section, we introduce the relevant signal processing notations and conventions  used to develop the analytical framework of this work. Some of these notations differ from those commonly used in digital signal processing(DSP). Table \ref{tab:table1} summarizes the notations adopted in this work, its standard DSP counterparts, and the relationship between them. These notations are intended to improve readability, ensure consistency throughout the manuscript, and  avoid ambiguity.

Continuous time analog signals are denoted using parentheses, such as $x(t)$, whereas discrete-time digital signals are represented using square brackets, as in $x[t_n]$. In the continuous case, t can assume any real value. In the discrete case, however, $t_n$ is restricted to the sampled time instants $t_n=n.T_s$, where $T_s$ is the sampling interval, and $n$ is an integer. The sampling rate is denoted as $F_s=1/T_s$. Further details can be found in the Supplementary Materials.

The frequency domain representation of $x(t)$, obtained via its Fourier transform, is denoted by $X(f)$. There are two common frequency domain representations of the discrete signal $x[t_n]$. One is the discrete time Fourier transform (DTFT) which is denoted as $\widetilde{X}(f)$ where $f$ is continuous. The other is the discrete Fourier transform (DFT) which is denoted as $\widetilde{X}[f_k]$, where $f_k=kF_s/N$ is a discrete variable, $k $ is an integer, and $N$ is the number of data points in the time series. While DTFT is applicable to discrete-time signals of both finite and infinite durations, DFT is applicable only to finite duration signals. 

The noise in the analog data is denoted by $\delta x(t)$. The power spectral density (PSD) of the analog noise is denoted as $S_{\delta x}(f)$. The noise in digitized data is denoted by $\delta x[t_n]$. Depending on the type of transform used, we denote the PSD as $\widetilde{S}_{\delta x}(f)$ - for DTFT, $\widetilde{S}_{\delta x}[f_k]$ - for DFT. Both $\widetilde{S}_{\delta x}(f)$ and $\widetilde{S}_{\delta x}[f_k]$ are obtained by dividing their respective standard forms by $F_s$, as summarized in Table \ref{tab:table1}. This ensures that they have the same dimensions as $S_{\delta x}(f)$ - the PSD of the analog noise.

Thus, parentheses and square brackets indicate whether the quantity enclosed in them is continuous or discrete, respectively. A tilde above the notation signifies a frequency domain representation of discrete time data, while its absence corresponds to a frequency domain representation of continuous time data. Throughout this work, the $sinc$ function is defined according to the standard DSP convention - $sinc(x)=\frac{sin(\pi x)}{\pi x}$.

\section{Experimental Methods }\label{sec:methods}

\begin{figure*}
 \includegraphics[]{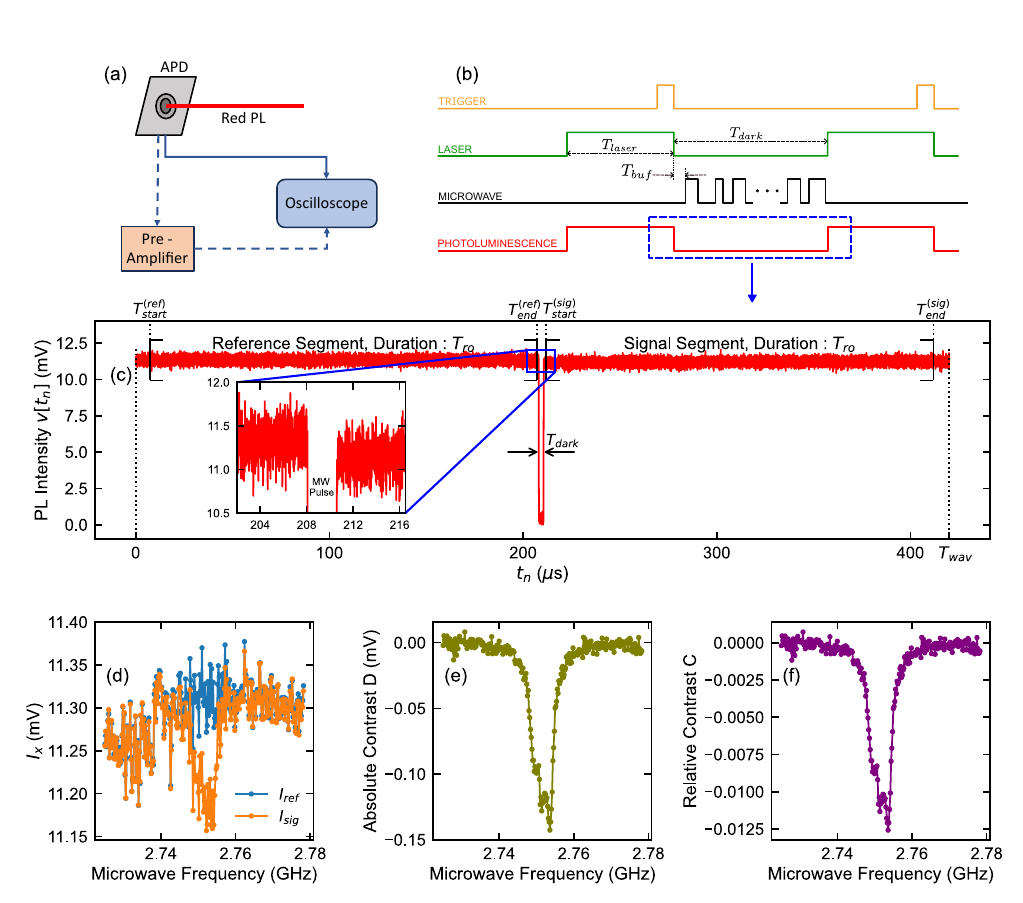}
\caption{\label{fig:methods}\textbf{(a)} Schematic of the detection part of our experimental setup. The red photoluminescence (PL) is collected by an avalanche photodetector (APD), and the output voltage is recorded either directly by the oscilloscope (solid blue line) or through a preamplifier (dashed blue line). \textbf{(b)} A representative pulse sequence for pulsed ODMR acquisition. The green pulse denotes the laser pulse sequence, with $T_{laser}$ and  $T_{dark}$ showing the ON and OFF duration of the laser pulse. The black pulse corresponds to a representative microwave (MW) sequence. A buffer interval $T_{buf}$ is introduced between the laser and MW pulses to allow completion of the optical cycle. The red pulse shows the resulting PL signal, and the orange pulse is used to trigger the oscilloscope acquisition.  \textbf{(c)} A waveform of duration $T_{wav}$, acquired in the oscilloscope by averaging over 32 repetitions of the pulse sequence. The signal and reference segments, each of duration $T_{\mathrm{ro}}$, are delineated by square brackets. The start and end of the reference segment are marked by $T_{start}^{(ref)}$ and $T_{end}^{(ref)}$ respectively, while $T_{start}^{(sig)}$ and $T_{end}^{(sig)}$ mark the corresponding boundaries of the signal segment. The inset highlights the MW-induced reduction in PL. \textbf{(d)} Variation of the $I_{\mathrm{ref}}$ and $I_{\mathrm{sig}}$ with MW frequency, shown in blue and orange, respectively.  \textbf{(e)},\textbf{(f)} The variation of the absolute contrast (D) and relative contrast (C), respectively, as a function of MW frequency.}
\end{figure*}

The electron spin dynamics in the triplet ground state manifold of the NV center is investigated with the help of pulsed optically detected magnetic resonance (ODMR). A detailed description of the defect system, and its energy level structure is beyond the scope of this work, and can be found elsewhere\cite{sewani2020coherent}. In the following, we give an operational description of the NV center and ODMR. There is a zero field splitting between the $m_s=0$ and $m_s=\pm 1$ sublevels of the triplet. A static magnetic field causes Zeeman splitting between $m_s=1$ and $m_s=-1$ sublevels. The transitions from $m_s=0$ to $m_s=-1$ or $m_s=1$ lie in the microwave frequency range. Thus, we can use microwaves, resonant with these transitions, to manipulate the spin state of the system. The changes in the spin population can be probed optically. The photoluminescence (PL) obtained on the application of an appropriate off-resonant (blue-shifted) laser excitation, is spin dependent, with a higher PL intensity corresponding to a larger population in the $m_s=0$ sublevel. We use a $532 nm$ green laser for excitation in order to maintain good charge stability\cite{aslam2013photo}. The photoluminescence  lies in the red region of the visible spectrum with a zero phonon line at $637 nm$.

The spin transitions correspond to energy splittings on the order of millikelvin. As a result, the thermal polarization is negligibly small at room temperature. The application of the off-resonant green laser drives a spin selective optical cycle which preferentially transfers population to the $m_s=0$ sublevel. Thus, the green laser performs spin polarization and readout, while the microwaves manipulate the spin state.

The full details of the experimental setup are provided in the Supplementary Materials. In this section, we describe the acquisition of the raw data using the oscilloscope, and also the data processing to get the final pulsed ODMR data.

\subsection{Data Acquisition}

A schematic of the detection part of the setup is shown in Figure \ref{fig:methods}(a). The red PL obtained from the sample is focused onto the Avalanche Photodetector (APD) which converts the optical intensity into analog voltage. This voltage is fed to the oscilloscope directly or through a preamplifier. A representative waveform obtained on the oscilloscope is shown in Figure \ref{fig:methods}(c).

Three detection configurations of the detection system, denoted as configurations 1, 2 and 3, are examined in this work. In configurations 1 and 2, the analog output voltage from the photodetector is fed directly to the oscilloscope. 
The analog bandwidth of the oscilloscope is set to $100 MHz$ in configuration 1, and $20 MHz$ in configuration 2. In configuration 3, the photodetector output is first passed through a preamplifier set at unity gain with a bandwidth of $1 MHz$, before being digitized by the oscilloscope, whose analog bandwidth is set to $20 MHz$. In this configuration, the preamplifier effectively serves as an analog low pass filter.

In a typical pulsed ODMR based protocol, non-overlapping laser and microwave pulses are applied, as illustrated in Figure \ref{fig:methods}(b). A green laser pulse of duration $T_{laser}$ - long enough to polarize the NV center spins - is applied. The laser is off for a duration $T_{dark}$ during which a microwave pulse sequence is delivered. A gap of $T_{buf}=1\mu s$ is left between the laser pulse and the microwave (MW) pulse in order to enable completion of the optical cycle, thus ensuring maximum polarization of the spins. An arbitrary MW pulse sequence is shown in Figure \ref{fig:methods}(b). However, the particular pulse sequence to be used depends on the protocol to be implemented, as discussed in Supplementary Materials. A second laser pulse is applied after the microwave pulse to readout the spin state and initialize the spins for the next pulse iteration.

Whenever the green laser pulse is on, red photoluminescence (PL)  is generated in the sample. The region of the PL shown in the dotted box of Figure \ref{fig:methods}(b) is captured in the oscilloscope time trace of Figure \ref{fig:methods}(c), and is denoted by $v[t_n]$. The waveform is acquired by averaging over $32$ repetitions of the pulse sequence, using the built-in waveform averaging function of the oscilloscope. This considerably reduces the acquisition time and minimizes the volume of data transferred from the oscilloscope. The oscilloscope needs to be triggered correctly, in order to observe a stable waveform on the screen, and ensure proper waveform averaging. The trigger is chosen in such a way that it has the same periodicity as one unit cell of the laser pulse, comprising $T_{laser}$ and $T_{dark}$. An example trigger signal is shown in Figure \ref{fig:methods}(b).

The acquired waveform  captures the PL towards the end of the first laser pulse and at the beginning of the second. The PL intensity towards the end of the first laser pulse is relatively flat, and corresponds to the maximally polarized state of the spin ensemble. At the beginning of the second laser pulse, the PL intensity is lower, as shown in the inset of Figure \ref{fig:methods}(c). This reduction happens because of the microwave pulse which decreases the population in the $m_s=0$ spin sublevel. As the second laser pulse continues, optical pumping repolarizes the NV centers into the $m_s=0$ state, leading to a gradual recovery of the PL intensity.

\subsection{Data Processing}

In the following, we describe the process of obtaining  useful data from the PL waveform. A reference and a signal segment of equal duration is selected as shown in Figure \ref{fig:methods}(c). The time duration of each segment is referred to as the readout time $T_{ro}$. The mean value of the data points within the signal and the reference segments are denoted by $I_{sig}$ and $I_{ref}$ respectively. The absolute contrast $D$ is defined as the difference between the two.

\begin{equation}
    D=I_{sig}-I_{ref}
    \label{eq:abs_con}
\end{equation}

With this convention, $D$ is non-positive. Alternatively, the difference is sometimes defined with the opposite sign, such that the contrast is non-negative. The relative contrast $C$ is obtained on dividing $D$ by $I_{ref}$, as shown in equation \eqref{eq:rel_con}. This represents the fractional change in the mean PL, and is the more commonly used measure of contrast.

\begin{equation}
    C=\frac{D}{I_{ref}}
    \label{eq:rel_con}
\end{equation}

In pulsed ODMR based protocols, the PL contrast is monitored as a function of various experimental parameters. Here, we consider the acquisition of pulsed ODMR spectrum, wherein the carrier frequency of a microwave pulse is varied. The variation of $I_{sig}$ and $I_{ref}$ with the microwave frequency is shown in Figure \ref{fig:methods}(d). The variation of absolute(D) and relative(C) contrast with microwave frequency is shown in Figure \ref{fig:methods}(e) and (f) respectively. The spectral features are obscured by noise in $I_{sig}$. By comparison, both $C$ and $D$ exhibit significantly less noise, enabling them to capture the spectrum accurately and reveal the hyperfine lines. The underlying reasons for this improvement are discussed in Section \ref{sec:noise_filt}.

\section{Characterization of Signal and Noise}\label{sec:sig_noise}

\subsection{Signal in Time and Frequency Domain}

The ODMR signal is closely connected to the repolarization dynamics of the 
photoluminescence (PL). $I_{sig}$ is obtained from a segment of the repolarization curve, as shown in Section \ref{sec:methods} B. Therefore, in order to investigate the dynamics in greater detail, we record the complete repolarization transient over a duration of $10 ms$, as illustrated in Figure \ref{fig:sig_noise}(a).  This curve deviates from the simple exponential behaviour observed for single NV centers\cite{suter2017single,mrozek2015longitudinal}. This happens as the spatially varying intensity profile of the laser leads to different repolarization rates at different parts of the ensemble. In order to ensure complete repolarization of the ensemble, a laser pulse $T_{laser}= 6ms$ is chosen based on Figure \ref{fig:sig_noise}(a). While this duration is longer than those used in single-defect experiments, it is consistent with values used in some ensemble-based studies \cite{mrozek2015longitudinal,jarmola2012temperature,gorrini2021long,masuyama2018extending}.

The amplitude spectrum of the repolarization curve is shown in the inset of Figure \ref{fig:sig_noise}(a). The 3-dB bandwidth of the repolarization signal is found to be $233 Hz$, whereas the $90 \%$  energy bandwidth extends to $3.88 kHz$. Given that the APD has bandwidth of $50 MHz$, and the oscilloscope offers a maximum bandwidth of $500 MHz$, the detection of this signal is well within the capabilities of the current setup. For higher laser powers, the NV centers can repolarize at significantly faster rates, resulting in signal bandwidths as high as $10 MHz$. Our detection system will perform reliably even under these conditions.

\subsection{Time Domain Noise in the Raw Data}

We now turn to a quantitative characterization of the noise present in the acquired data. Previous studies have identified laser power fluctuations and the photodetector noise as the dominant contributors\cite{zhang2019modified}. In the present work, we do not resolve the individual contributions but instead analyze the combined noise.

In Section \ref{sec:methods}, the sampled data within an individual waveform has been denoted by $v[t]$. In order to distinguish between waveforms, we now adopt the notation $v_{t'}[t_n]$, where $t'$ denotes the instant at which the trigger corresponding to a given waveform is detected. While $t_n$ is the local time coordinate within an acquired waveform, $t'$ labels successive waveform acquisitions.

The acquired data can be regarded as the sum of an ideal noiseless component, $v^{*}_{t'}[t_n]$, and a noise-induced deviation, $\delta v_{t'}[t_n]$, as expressed in equation \eqref{eq:sig_noise}.

\begin{equation}
    v_{t'}[t_n] = v_{t'}^{*}[t_n]+\delta v_{t'}[t_n]
    \label{eq:sig_noise}
\end{equation}

The noise can be further separated into two components, as shown in equation \eqref{eq:noise_comp}. The first component corresponds to fluctuations within a single waveform, and is denoted by $\delta v^{(fl)}_{t'}[t_n]$. The second component does not produce temporal fluctuations within a waveform; instead it appears as a waveform-dependent offset that varies between acquisitions. This quasi-static shift component is denoted by $\delta v^{(qs)}_{t'}$.

\begin{equation}
    \delta v_{t'}[t_n]=\delta v^{(fl)}_{t'}[t_n]+\delta v^{(qs)}_{t'}
    \label{eq:noise_comp}
\end{equation}

The intra-waveform noise is quantified by the standard deviation of the fluctuations $\delta v^{(fl)}_{t'}[t_n]$, and is denoted as $\sigma_{raw}$. However, the inter-waveform noise includes significant drift, and is not reliably captured by the standard deviation. To quantify this noise, we use the Allan deviation, denoted as $\sigma^{(A)}_{raw}(\tau)$, which measures the magnitude of variations in $\delta v^{(qs)}_{t'}$ over a timescale $\tau$. The subscript `raw' indicates that these quantities are evaluated using the unprocessed waveform data, and is used to distinguish them from the corresponding metrics calculated after data processing. The procedures used to calculate the standard deviation and the Allan deviation are described in Appendix \ref{app:sd_ad}.

\begin{figure}
 \includegraphics[]{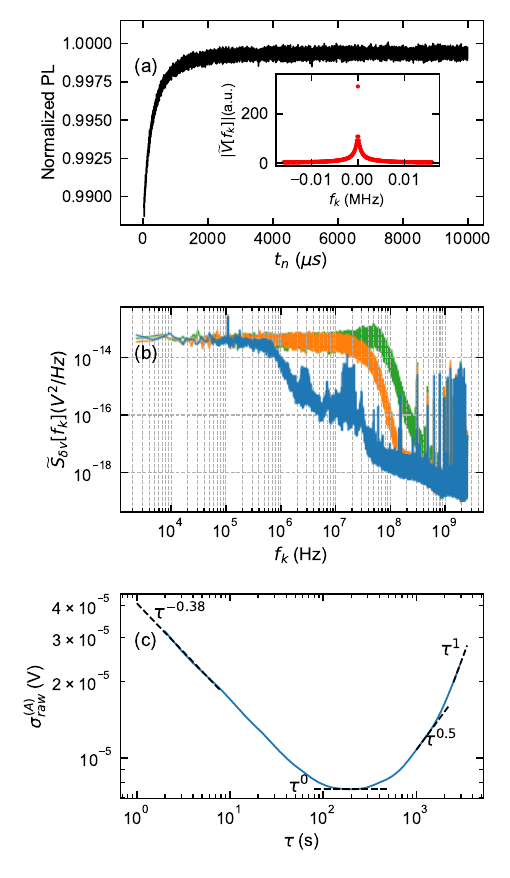}
 \centering
\caption{\label{fig:sig_noise}\textbf{(a)} Variation of the red PL with time during the repolarization process; the inset displays the corresponding amplitude spectrum.
\textbf{(b)} Power spectral density (PSD) for detection configurations 1 (green), 2 (orange) and 3 (blue). \textbf{(c)} The variation of Allan deviation with the timescale $\tau$ over which the changes in inter-waveform noise are captured. The black dashed lines indicate the local logarithmic slope of the Allan deviation. 
}
\end{figure}

The measured Allan deviation, depicted in Figure \ref{fig:sig_noise}(c), reveals that different noise processes dominate at different timescales. If the Allan deviation scales as $\tau^{\alpha}$ over a given range of $\tau$, then the exponent $\alpha$, which is also the slope in the log-log graph, directly indicates the underlying noise mechanism \cite{IEEE1139}. For lower values of $\tau$, from a few seconds to a couple of minutes, the Allan deviation decreases with increasing $\tau$ with $-0.5<\alpha<0$. This behavior indicates a combination of white noise ($\alpha=-0.5$) and flicker noise($\alpha=0$). As $\tau$ increases, the flicker noise contribution becomes dominant, marked by the flattening of the Allan deviation ($\alpha \approx 0$), occurring in the vicinity of a few minutes. For larger $\tau$, the Allan deviation turns upwards, indicating the onset of drift-dominated behavior, with $0<\alpha\leq1$. In this regime, $\alpha \approx 0.5$ signifies random drift, while $\alpha \approx 1$ indicates linear systematic drifts.

\subsection{Frequency Domain Noise in the Raw Data}
The power spectral density (PSD) of the noise is estimated from the acquired waveform data using the Average Periodogram method, as described in the Appendix \ref{app:psd_est}. The resulting discrete PSD, $\widetilde{S}_{\delta v}[f_k]$, is shown in Figure \ref{fig:sig_noise}(b), with adjacent frequency points separated by $\Delta f=1/T_{wav}$. Conceptually, the noise can be associated with an underlying continuous power spectral density, denoted $\widetilde{S}_{\delta v}(f)$, corresponding to the limit of an infinitely long acquisition. The experimentally obtained PSD  $\widetilde{S}_{\delta v}[f_k]$ is related to the underlying spectrum $\widetilde{S}_{\delta v}(f)$ via equation \eqref{eq:spectral_leakage}.

\begin{equation}
    \widetilde{S}_{\delta v}[f_k]
    =
    \left.
    \left(
    \widetilde{S}_{\delta v}(f)
    \circledast
    \widetilde{W}(f)
    \right)
    \right|_{f=f_k}
    \label{eq:spectral_leakage}
\end{equation}

Here, $\widetilde{W}(f)$ is the DTFT of the triangular Bartlett function, and $\circledast$ denotes periodic convolution\cite{proakis2007digital}. The convolution with $\widetilde{W}(f)$ spreads the spectral content over neighboring frequencies, a phenomenon commonly referred to as spectral leakage. Equation \eqref{eq:spectral_leakage} along with the exact form of $\widetilde{W}(f)$ have been derived in Appendix \ref{app:spec_leak}.

The PSDs corresponding to the digitized data - $\widetilde{S}_{\delta v}[f_k]$ and $\widetilde{S}_{\delta v}(f)$ - are related to, but not identical to, $S_{\delta v}(f)$, the PSD of the noise in the analog voltage at the oscilloscope input. The digitization process introduces both quantization noise and spectral aliasing. As shown in Appendix \ref{app:quant_noise}, the quantization noise arising from the finite resolution of the analog to digital converter (ADC) is negligible compared to the optical noise. However, the finite sampling rate $F_s$ limits the observable frequency range to the Nyquist frequency, $F_s/2$. As a result, frequency components of the noise outside this range are folded back into the observable spectrum,  giving rise to aliases, as described by equation \eqref{eq:aliasing}\cite{Kirchner2005}. This phenomenon, known as aliasing\cite{proakis2007digital}, has significant implications for pulsed ODMR measurements, as described in Section \ref{sec:samp_rate}.

\begin{equation}\label{eq:aliasing}
    \widetilde{S}_{\delta v}(f)=\sum_{k=-\infty}^{+\infty} S_{\delta v}(f-kF_s)
\end{equation}

Figure \ref{fig:sig_noise}(b) shows the power spectral density for configurations 1, 2 and 3 of the detection setup in green, orange and blue respectively. Configuration 1 captures the complete noise spectral density because its analog acquisition bandwidth is higher than the $50 MHz$ analog bandwidth of the avalanche photodetector. We therefore first examine the green curve corresponding to configuration 1. At lower frequencies, the PSD remains approximately flat at a level of order $10^{-14}$ $V^2/Hz$. As the frequency increases, the PSD rises, and exhibits a pronounced peak close to the APD bandwidth. Such high frequency peaking is characteristic of trans-impedance amplifier based photodetectors \cite{graeme1995photodiode}, and is also consistent with the device specifications.
 Beyond the APD bandwidth, the spectral components of the noise are progressively attenuated, leading to monotonic decrease of the PSD with frequency. The PSD of configurations 2 and 3 closely follow the PSD of configuration 1 within their respective analog bandwidths. Beyond these frequencies, the noise spectral components are attenuated.

\subsection{Connecting the Temporal and Spectral Measures of Noise}

The standard deviation of the fluctuations within the waveform quantified by the $\sigma_{raw}$ is connected to the power spectral density by Parseval's theorem according to equation \eqref{eq:sigma_raw} where $M$ is the number of samples in the recorded waveform. 

\begin{equation}
    \sigma^2_{raw}  \approx 2 \sum_{n=1}^{\left\lfloor (M-1)/2 \right\rfloor} \widetilde{S}_{\delta v}\Bigl[\frac{n}{T_{wav}}\Bigr]\cdot \frac{1}{T_{wav}} 
    \label{eq:sigma_raw}
\end{equation}

The $\sigma_{raw}$ is connected to the continuous PSD via equation \eqref{eq:sigma_raw_cont_psd}. The relation follows from equation \eqref{eq:sigma_raw} and the spectral leakage relation of equation \eqref{eq:spectral_leakage} as shown in Appendix \ref{app:spec_cont}.

\begin{equation}
    \sigma^2_{raw} \approx 2\int_{0}^{F_s/2}  \widetilde{S}_{\delta v}(f) \widetilde{W}'(f) df
    \label{eq:sigma_raw_cont_psd}
\end{equation}

The Allan deviation $\sigma_{raw}^{(A)}(\tau)$, which captures the inter-waveform variations is connected to the power spectral density according to equation \eqref{allan_spectrum_connection}. 

\begin{equation}
    (\sigma_{raw}^{(A)}(\tau))^2  \approx \int_{0}^{\infty} \widetilde{S}_{\delta v}(f) \widetilde{W}''(f)df
    \label{allan_spectrum_connection}
\end{equation}

Thus, the functions $\widetilde{W}'(f)$ and $\widetilde{W}''(f)$ determine which spectral components contribute to the standard deviation and Allan deviation, respectively. The analytical form of these functions, along with their plots, have been presented, and discussed in detail in Appendix \ref{app:spec_cont}. The analysis shows that the standard deviation is very strongly affected by spectral components above $1/T_{wav}$, while the Allan deviation is primarily sensitive to spectral components below $1/2T_{wav}$. Spectral components between $1/T_{wav}$ and $1/2T_{wav}$ contribute to both the standard and Allan deviation. These results establish that the standard deviation is primarily influenced by high frequency fluctuations while the Allan deviation is more sensitive to the low frequency components of the noise.

\subsection{Noise in the ODMR Data and Contrast to Noise Ratio}

The noise in the processed ODMR data, represented by the absolute contrast $D$ or the relative contrast $C$, can be fully characterized by the standard deviation alone, as will be shown in Section \ref{sec:noise_filt}. The corresponding standard deviations are denoted as $\sigma_{odmr}^{(D)}$ and $\sigma_{odmr}^{(C)}$ respectively.

While $\sigma_{odmr}$ provides useful information, it does not by itself determine the quality of the processed ODMR data. A more appropriate figure of merit (FOM) is the contrast to noise ratio (CNR), which accounts for both signal contrast and noise. Let the maximum contrast in the ODMR spectrum be denoted as $D_{max}$ or $C_{max}$, depending on the contrast metric used. The contrast to noise ratio (CNR) is then defined by equation \eqref{eq:cnr_general}, where $X$ represents either $D$ or $C$.

\begin{equation}
    CNR^{(X)} = \frac{X_{max}}{\sigma_{odmr}^{(X)}}
    \label{eq:cnr_general}
\end{equation}

The method for calculating the CNR from the ODMR spectrum is described in the Appendix \ref{app:calc_cnr}. As shown in the Appendix \ref{app:comp_cnr}, for all the cases considered in this work, the CNRs corresponding to $C$ and $D$ are approximately equal. Since the relative contrast $C$ is the more widely used measure, the contrast, noise and CNR presented throughout this paper correspond to $C$. For brevity, the superscript will be omitted hereafter, with the understanding that all quantities refer to the relative contrast $C$ unless stated otherwise.

\section{Data Processing as a Digital Filter}\label{sec:noise_filt}

\begin{figure*}
\includegraphics[]{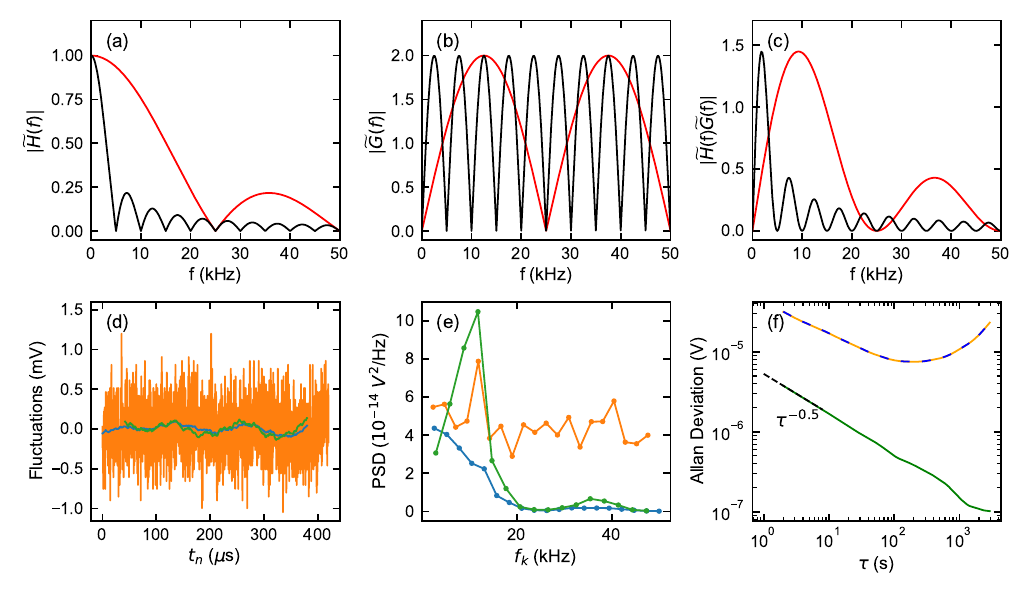}
\caption{\label{fig:noise_filter}\textbf{(a-c)} The red and black curves correspond to readout durations $T_{ro}=40 \mu s$ and $100 \mu s$, respectively. \textbf{(a)} Amplitude response of the filter function $\widetilde{H}(f)$ corresponding to time-averaging over the readout duration $T_{\mathrm{ro}}$. It exhibits a low-pass response. 
\textbf{(b)} Amplitude response of the filter function $\widetilde{G}(f)$ associated with the subtraction operation. It exhibits a low-reject (high-pass) response.
\textbf{(c)} Amplitude response of the effective filter $\widetilde{H}(f)\widetilde{G}(f)$ which models the ODMR data processing. It is a band-pass filter.
\textbf{(d)} The intra-waveform fluctuations in the raw data $v[t_n]$ (orange). The fluctuations in the time series $\overline{v}[t_n]$, obtained by performing time-averaging on the raw data (blue). The fluctuations in $d[t]$, obtained by performing differencing operations on the time-averaged data (green). 
\textbf{(e)} The PSD corresponding to $v[t_n]$ (orange), $\overline{v}[t_n]$ (blue) and $d[t_n]$ (green). \textbf{(f)} The Allan deviation of the inter-waveform noise in $v[t_n]$ (orange), $\overline{v}[t_n]$ (blue) and $d[t_n]$ (green). The black dotted line represents the local slope of the Allan deviation.}

\end{figure*}

The impact of the data processing described in Section \ref{sec:methods} on the noise in the final ODMR data, can be interpreted as the action of an effective filter implicitly implemented by the processing algorithm. In this section, we derive the transfer function of the filter and show how it reduces the noise in the processed ODMR data by suppressing noise over a broad range of frequencies present in the raw data. We identify the spectral band which is only weakly attenuated by the filter, and therefore provides dominant contribution to the noise in the final data.

\subsection{Filter Model for Segment Mean}

The quantities $I_{sig}$ and $I_{ref}$ were obtained by taking the mean of the signal and reference segments respectively, as described in Section \ref{sec:methods}. To examine the effect of this averaging operation, a new time series $\overline{v}[t_n]$ is defined. It is derived from the raw waveform data $v[t_n]$, by replacing each data point with the mean of the preceding data over a duration $T_{ro}$. This operation is defined mathematically in equation  \eqref{eq:vbar}.

\begin{equation}
    \overline{v}[t_n]=\frac{1}{N_{ro}}\sum_{k=n-N_{ro}+1}^{n} v[t_k]
    \label{eq:vbar}
\end{equation}

Here, $N_{ro}$ is the number of samples within the duration $T_{ro}$, such that $(N_{ro}-1)T_s=T_{ro}$. It follows from equation \eqref{eq:vbar} and Figure \ref{fig:methods}(c) that $I_{sig}$ and $I_{ref}$ can be expressed by equation \eqref{eq:vbar_isig_iref}, where $T^{(sig)}_{end}$ and $T^{(ref)}_{end}$ are the end-points of the signal and reference segments, respectively.

\begin{equation}
    I_{\mathrm{sig}}=\overline{v}[T^{(\mathrm{sig})}_{\mathrm{end}}],
    \qquad
    I_{\mathrm{ref}}=\overline{v}[T^{(\mathrm{ref})}_{\mathrm{end}}]
    \label{eq:vbar_isig_iref}
\end{equation}

This shows that both quantities correspond to specific samples of $\overline{v}[t_n]$. Thus, the origin of the noise observed in the $I_{sig}$ and $I_{ref}$ curves in Figure \ref{fig:methods}(d) can be understood from  the noise properties of $\overline{v}[t_n]$.  
 
Equation \eqref{eq:vbar} clearly has the form of a finite impulse response (FIR) digital filter \cite{proakis2007digital}. $\overline{v}[t_n]$ can be written as a convolution, $\overline{v}[t_n]=h[t_n]*v[t_n]$, where the impulse response $h[t_n]$ is a rectangular function as defined in equation \eqref{h_small} with $u[t_n]$ being the Heaviside step function.

\begin{equation}
    h[t_n]=\frac{1}{N_{ro}}u[t_n]u[T_{ro}-t_n]
    \label{h_small}
\end{equation}

The frequency response $\widetilde{H}(f)$ of the filter is obtained by the DTFT of $h[t_n]$. It is found to be related to the Dirichlet kernel, as shown in equation \eqref{eq:dirichlet}. 

\begin{equation}
    \widetilde{H}(f)=e^{-i\pi fT_{ro}}\frac{sinc( f (T_{ro}+T_{s}))}{sinc( f T_{s})}
    \label{eq:dirichlet}
\end{equation}

Only frequency components of the filter till the Nyquist frequency ($f=F_s/2$) influence the data; contributions beyond that can be safely ignored\cite{proakis2007digital}. Within this relevant frequency range, $\widetilde{H}(f)$ can be approximated by equation \eqref{h_big} for $T_{ro}>>T_s$, which holds for all the experimental conditions considered in this work.

\begin{equation}
     \widetilde{H}(f) \approx e^{-i\pi fT_{ro}}sinc( fT_{ro})
     \label{h_big}
 \end{equation}

The amplitude response of the filter is shown in Figure \ref{fig:noise_filter}(a), where the red and black curves correspond to $T_{ro}=40 \mu s$ and $T_{ro}=100 \mu s$ respectively. It is found that increasing $T_{ro}$ leads to a narrower passband, and a correspondingly sharper attenuation of higher-frequency components. The color convention described here applies to Figure \ref{fig:noise_filter}(b) and (c) also.

The impact of the filter on the raw data is illustrated in Figure \ref{fig:noise_filter}(d). The raw data $v[t_n]$ obtained from the oscilloscope, shown in orange, is extremely noisy. The time-series $\overline{v}[t_n]$, obtained with $T_{ro}=40 \mu s$ and shown in blue, exhibits markedly reduced fluctuations. This improvement arises from the attenuation of the higher frequency components of the PSD, which dominate the intra-waveform noise, as discussed in Subsection C of Section \ref{sec:sig_noise}. The experimentally obtained PSDs corresponding to $v[t_n]$ and $\overline{v}[t_n]$, shown in orange and blue respectively in Figure \ref{fig:noise_filter}(e), further confirms the low pass filtering action of the averaging operation. The Allan deviation, on the other hand, remains unchanged relative to that of the raw data, as seen in Figure \ref{fig:noise_filter}(f). The orange curve represents the Allan deviation of $v[t_n]$, while the blue dashed curve corresponds to that of $\overline{v}[t_n]$; the two overlap completely. This is expected since the low frequency components which dominate Allan deviation is largely preserved by the low pass nature of this filter. The high Allan deviation results in substantial inter-waveform noise, leading to pronounced fluctuations and drift in $I_{sig}$ and $I_{ref}$, as seen in Figure \ref{fig:methods}(d).

\subsection{Filter Model for Reference Subtraction}

The absolute contrast $D$ is obtained by subtracting $I_{ref}$ from $I_{sig}$, as given in equation \eqref{eq:abs_con}. In order to analyze the effect of this differencing operation, we introduce a time-series $d[t_n]$, defined in equation \eqref{eq:d_small}. By construction, the absolute contrast satisfies $D=d[T^{(sig)}_{end}]$. This follows from equations \eqref{eq:abs_con},\eqref{eq:vbar_isig_iref} and Figure \ref{fig:methods}(c). Similar to the previous analysis, this formulation allows us to infer the noise characteristics of $D$ by examining the noise properties of $d[t_n]$.

\begin{equation}
    d[t_n]=\overline{v}[t_n]-\overline{v}[t_n-T_{ro}-T_{dark}]
    \label{eq:d_small}
\end{equation}

Equation \eqref{eq:d_small} also represents a digital filter operation, and can be written as $d[t_n]=g[t_n]*\overline{v}[t_n]$ where $g[t_n]$ is the impulse response of the filter. In the present case, $g[t_n]$ is given by the difference of two delta functions as expressed in equation \eqref{g_small}.

\begin{equation}
    g[t_n]=\delta[t_n]-\delta[t_n-T_{ro}-T_{dark}] 
    \label{g_small}
\end{equation}

The impulse response $\widetilde{G}(f)$, obtained from the DTFT of $g[t_n]$, is a sinusoidal function, as expressed in equation \eqref{big_g}.

\begin{equation}
    \widetilde{G}(f)=2i e^{-i \pi f (T_{ro}+T_{dark})} sin(\pi f (T_{ro}+T_{dark}))
    \label{big_g}
\end{equation}

 The condition $T_{dark}<<T_{ro}$ holds across all values of $T_{ro}$ considered in this work. This allows us to approximate $\widetilde{G}(f)$ by equation \eqref{big_g_approx}. We emphasize that this approximation does not hold for all other pulsed ODMR based protocols, as can be seen in the Supplementary Materials.

\begin{equation}
    \widetilde{G}(f)\approx 2i e^{-i \pi f T_{ro}} sin(\pi f T_{ro})
    \label{big_g_approx}
\end{equation}

The amplitude response of this filter, shown in Figure \ref{fig:noise_filter}(b), clearly demonstrates its high pass character with strong attenuation of the low frequency components. The extent of this attenuation is dependent on the readout time $T_{ro}$ : increasing $T_{ro}$ reduces the suppression of the low frequency components.

The time series $d[t_n]$ is shown by the green curve in Figure \ref{fig:noise_filter}(d). Its fluctuations are comparable to those of $\overline{v}[t_n]$, shown in blue, indicating that the subtraction step does not substantially alter the intra-waveform noise. On the other hand, the Allan deviation corresponding to $d[t_n]$, shown by the green curve in Figure \ref{fig:noise_filter}(f), is substantially smaller than those of $\overline{v}[t_n]$ and $v[t_n]$. This is expected since lower frequency components which contribute most strongly to the Allan deviation are filtered out. This reduction in the Allan deviation indicates that the quasi-static inter-waveform noise is effectively suppressed due to the differencing operation. 

The filtering action of the subtraction substantially reduces noise levels in the absolute contrast $D$, compared with that in $I_{sig}$ and $I_{ref}$, as evident from Figures \ref{fig:methods} (d) and (e). The fluctuations and drifts in $I_{ref}$ and $I_{sig}$ are highly correlated, as both are influenced by the low frequency quasi-static inter-waveform noise. Subtracting these signals therefore suppresses this common-mode noise, thereby improving the data quality of the absolute contrast $D$.

\subsection{Effective Filter Model for Data Processing}

The processing of the raw waveform data to obtain the ODMR data, given by the absolute contrast $D$, is modeled by the effective filter which transforms $v[t_n]$ into $d[t_n]$.
This filter is characterized by the transfer function $\widetilde{H}(f) \widetilde{G}(f)$ - the product of the transfer functions of the mean and difference filters. The corresponding amplitude response, shown in Figure \ref{fig:noise_filter}(c), exhibits a bandpass characteristic. This is further confirmed from the PSD curve corresponding to $d[t_n]$, shown in Figure \ref{fig:noise_filter}(e). As shown in Figure \ref{fig:noise_filter}(f), throughout the measured range of $\tau$, the Allan deviation corresponding to $d[t_n]$ scales approximately as $\tau^{-0.5}$, indicative of predominantly white noise. Hence, for ODMR data represented by the absolute contrast $D$, the noise can be adequately characterized by its standard deviation alone, denoted by $\sigma_{odmr}^{(D)}$. This quantity is related to the noise power spectral density through the filter transfer function, as described by equation \eqref{eq:sigma_odmr_psd}.

\begin{equation}
    (\sigma^{(D)}_{odmr})^2 \approx 2\int_{0}^{F_s/2}  \widetilde{S}_{\delta v}(f) |\widetilde{H}(f) \widetilde{G}(f)|^2 df
    \label{eq:sigma_odmr_psd}
\end{equation}

Notably, equation \eqref{eq:sigma_odmr_psd} strictly holds when the data are acquired over only one iteration of the pulse sequence for each microwave frequency. When averaging over several cycles of the pulse sequence, the PSD $\widetilde{S}_{\delta v}(f)$ is replaced by that of the averaged waveform, as detailed in Section \ref{sec:wav_avg}. However, the conclusions drawn from this analysis remain valid.

The amplitude spectrum of the filter, shown in Figure \ref{fig:noise_filter}(c), indicates that the primary contribution to $\sigma_{odmr}$ arises from an intermediate frequency (IF) band, since fluctuations at very low and very high frequencies are strongly suppressed. As quantified in Appendix \ref{app:if_band}, the IF band spans from $F_l$ to $F_u$, with a width $F_{if}=F_u-F_l$. Its extent is governed by the readout time $T_{ro}$, with longer readout time producing a  narrower IF band, as illustrated in Figure \ref{fig:noise_filter}(c). Therefore, for a given $T_{ro}$, improving the contrast to noise ratio requires suppressing the spectral contributions of noise specifically from this IF band. Unless stated otherwise, $T_{ro}=200 \mu s$ is used throughout this work. For this $T_{ro}$, the IF band, as defined here, extends from $0.12 kHz$ to $18.2 kHz$.

The digital filter formalism described in this section is applicable only to the absolute contrast $D$. However, if some conclusion is reached about $CNR^{(D)}$ using this formalism, it will also hold for $CNR^{(C)}$.  This is because under the experimental conditions considered in this work, the two quantities are approximately equal, as given in equation \eqref{eq:cnr_approx_equal} and shown in Appendix \ref{app:comp_cnr}. 

\begin{equation}
    \frac{C_{max}}{\sigma_{odmr}^{(C)}} \approx \frac{D_{max}}{\sigma_{odmr}^{(D)}}
    \label{eq:cnr_approx_equal}
\end{equation}

\section{Waveform Averaging as a Digital Filter}\label{sec:wav_avg}

\begin{figure*}
\centering
\includegraphics[]{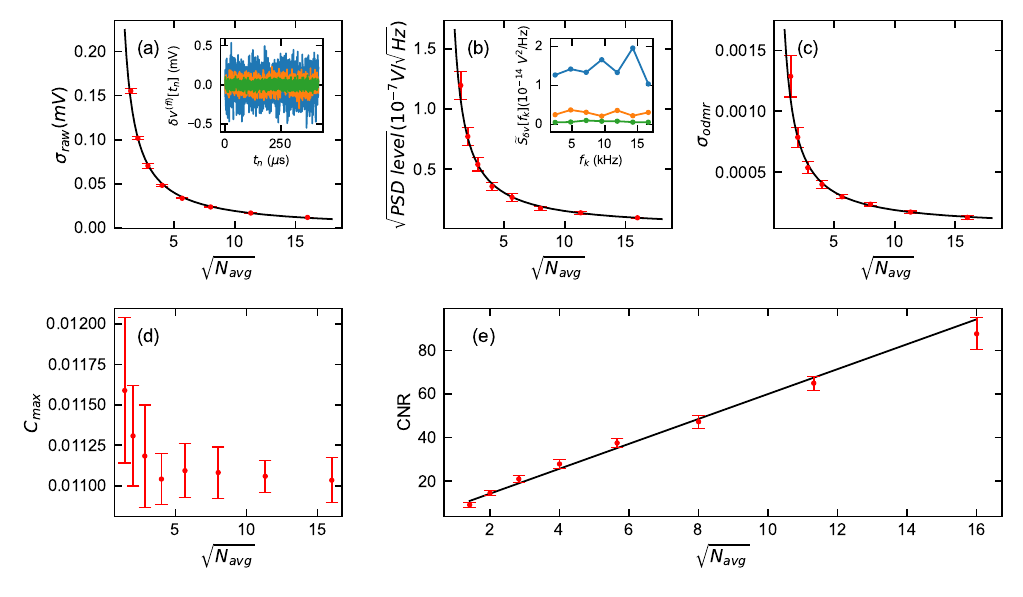 }
\caption{\label{fig:wav_avg} \textbf{(a-c)} Variation  of the noise in the raw data ($\sigma_{\mathrm{raw}}$), square root of the IF-band PSD level and the noise in the processed ODMR data($\sigma_{\mathrm{odmr}}$), respectively, with the square root of number of waveform averages,$\sqrt{N_{avg}}$. The black line in each panel shows a $1/\sqrt{N_{\mathrm{avg}}}$ fit. The inset of \textbf{(a)} shows representative intra-waveform noise in the raw data for $N_{\mathrm{avg}}=2$, 8 and 32 (blue, orange and green, respectively). The inset of \textbf{(b)} shows the corresponding PSD within the intermediate-frequency (IF) band using the same color scheme. \textbf{(d)} Variation of the ODMR contrast  as a function of $\sqrt{N_{avg}}$. \textbf{(e)} Variation of Contrast to Noise Ratio with $\sqrt{N_{avg}}$. The black curve is the linear fit to the data. All data were obtained in configuration 3, with $F_s=11.9 Msps$, $T_{ro}=200 \mu s$.}
\end{figure*}

In this section, we model the effect of  averaging multiple waveforms, each corresponding to an iteration of the pulse sequence, as a digital filter. Using this model, we investigate how the waveform averaging influences the contrast to noise ratio (CNR) of the processed data.

The relevant portion of the waveform is captured on the oscilloscope screen with the appropriate triggering as described in Section \ref{sec:methods}. In the present case, waveforms triggered at intervals of $T_{trig}=T_{laser}+T_{dark}$ are acquired, and averaged on the oscilloscope. Therefore, the data in the averaged waveform can be represented by equation \eqref{eq:wav_avg_sub}.


\begin{equation}
    v^{(avg)}[t_n]=\frac{1}{N_{avg}}\sum_{l=0}^{N_{avg}-1} v[t_n-l\cdot T_{trig}]
    \label{eq:wav_avg_sub}
\end{equation}

The impact of the averaging process on the temporal noise is shown in 
Figure \ref{fig:wav_avg}(a). The inset shows the fluctuations in the raw data for $N_{avg}=2$ (blue), $8$ (orange) and $32$ (green). It is evident that the extent of fluctuations in the raw data decreases with increasing $N_{avg}$. The dependence of the corresponding standard deviation, $\sigma_{raw}$, on $N_{avg}$ is shown in red in Figure \ref{fig:wav_avg}(a), while the black fit shows $1/\sqrt{N_{avg}}$.

We introduce a digital filter model to characterize the effect of waveform averaging on the spectral components of the noise. The digital filter corresponding to equation \eqref{eq:wav_avg_sub} has a transfer function $\widetilde{K}(f)$, given by equation \eqref{eq:k_tilda}. It can be obtained in a way identical to that of $\widetilde{H}(f)$, derived in Section \ref{sec:noise_filt}. 

\begin{equation}
    \widetilde{K}(f) = e^{-i \pi f(N_{avg}-1)T_{trig}}\frac{sinc( f N_{avg}T_{trig})}{sinc( f T_{trig})}
    \label{eq:k_tilda}
\end{equation}

Accordingly, the underlying power spectral density (PSD) of the averaged waveform - $\widetilde{S}_{\delta v^{(avg)}}(f)$ - gets modified according to equation \eqref{eq:psd_v_avg}.

\begin{equation}
    \widetilde{S}_{\delta v^{(avg)}}(f)=|\widetilde{K}(f)|^2 \widetilde{S}_{\delta v}(f)
    \label{eq:psd_v_avg}
\end{equation}

Hence, $\sigma^{(D)}_{odmr}$ still obeys equation \eqref{eq:sigma_odmr_psd} with  $\widetilde{S}_{\delta v}(f)$ replaced by $\widetilde{S}_{\delta v^{(avg)}}(f)$. Combining this result with the discussion in Section \ref{sec:noise_filt}, we conclude that the noise in the processed ODMR data is determined by the spectral components of $\widetilde{S}_{\delta v^{(avg)}}(f)$ within the intermediate frequency (IF) band. Since the experimentally estimated PSD,  $\widetilde{S}_{\delta v^{(avg)}}[f_k]$ is discrete, we examine its spectral content in this frequency range, as shown in Figure \ref{fig:wav_avg}(b).

The inset of Figure \ref{fig:wav_avg}(b) presents $\widetilde{S}_{\delta v^{(avg)}}[f_k]$ in the IF band for different number of averages - $N_{avg}=2$(blue), $8$(orange) and $32$(green). The characteristic Dirichlet kernel structure is not apparent because it is obscured by spectral leakage, as described in Appendix \ref{app:psd_wav_avg}. Instead, the discrete PSD is well approximated by equation \eqref{eq:psd_wav_approx}. Notably, this approximation holds even when the waveforms are triggered at unequal intervals, which may occur in practice.

\begin{equation}
    \widetilde{S}_{\delta v^{(avg)}}[f_k] \approx \frac{1}{N_{avg}} \widetilde{S}_{\delta v}[f_k]
    \label{eq:psd_wav_approx}
\end{equation}

To compare the noise for different values of $N_{avg}$, we define the IF PSD level to be the mean PSD over the IF band. Figure \ref{fig:wav_avg}(b) (red) shows the variation of the square root of the IF PSD level with $\sqrt{N_{avg}}$. The data follow the expected $1/\sqrt{N_{avg}}$ dependence, indicated by the black line. Consistent with this behavior, $\sigma_{odmr}$ follows a similar trend, as shown in Figure \ref{fig:wav_avg}(c). The ODMR contrast is approximately independent of $N_{avg}$, as shown in Figure \ref{fig:wav_avg}(d). The contrast to noise ratio (CNR) therefore scales linearly with $\sqrt{N_{avg}}$, as seen in Figure \ref{fig:wav_avg}(e). While increasing the number of averages improves CNR, it also increases the acquisition time, as shown in the Appendix \ref{app:acq_oh}.   

In this particular case, it is noteworthy that the fluctuations in the raw data ($\sigma_{raw})$ and the processed ODMR data ($\sigma_{odmr})$ exhibit similar behavior. However, in Section \ref{sec:samp_rate}, we show that this correspondence does not always hold.

\section{Impact of Readout Duration} \label{sec:readout}

\begin{figure}
\centering
\includegraphics[width=\linewidth]{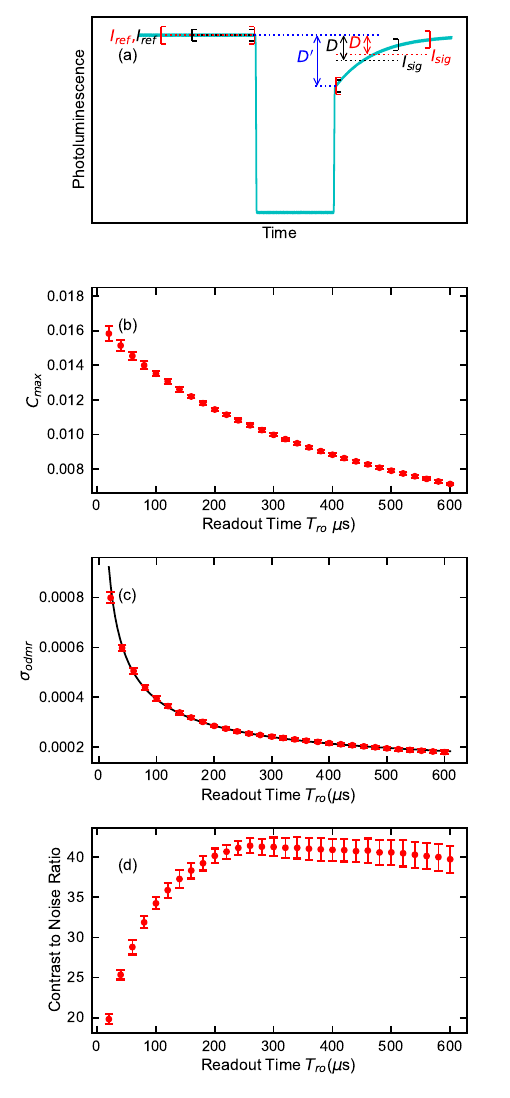}
\caption{\label{fig:ro_combined} \textbf{(a)} Schematic representation of the raw waveform data. The reference and signal segments are enclosed within square brackets, with the red brackets spanning a longer duration than the black ones. The voltage levels corresponding to $I_{sig}$ and $I_{ref}$ are indicated by dotted horizontal lines of the appropriate colour. $D'$ denotes the maximum contrast. The magnitude of the absolute contrast $D$ is indicated by double-headed arrows of the corresponding colours. 
\textbf{(b–d)} Variation of contrast, noise and contrast-to-noise ratio, respectively,  as functions of the readout duration $T_{ro}$. The black line in panel \textbf{(c)} shows a $1/\sqrt{T_{ro}}$ fit. All data were acquired in configuration 3, with $F_s=11.9Msps$, $N_{avg}=32$.
}
\end{figure}

This section examines the impact of the readout duration on the contrast to noise ratio (CNR) of the processed ODMR data. The results are summarized in Figure \ref{fig:ro_combined}. The observed variation in the contrast is interpreted through a time-domain analysis, while the corresponding variation in noise is explained using the frequency-domain filtering, introduced in Section \ref{sec:noise_filt}. Notably, drawing conclusions about the contrast variation from the frequency domain analysis is non-trivial, as discussed in the Appendix \ref{app:con_freq}.

Figure \ref{fig:ro_combined}(a) presents a schematic representation of the photoluminescence (PL) data shown in Figure \ref{fig:methods}(c). The reference and signal segments are indicated by square brackets, following the same convention used in Figure \ref{fig:methods}(c). Two distinct readout durations $T_{ro}$ are illustrated for comparison. The segments enclosed by the red brackets correspond to a longer readout duration compared to those enclosed by the black brackets. The voltage level corresponding to $I_{sig}$ and $I_{ref}$ are shown by dotted horizontal lines, with the colour denoting the associated readout duration. As previously described in Section \ref{sec:methods}, $I_{sig}$ and $I_{ref}$ are simply the mean values of the data points within the respective segments.

The schematic demonstrates that $I_{sig}$ increases with increasing readout duration, whereas $I_{ref}$ remains approximately constant, provided that $T_{ro}$ is sufficiently smaller than $T_{laser}$. Consequently, the magnitude of the absolute contrast $D$ decreases for longer readout durations. It follows from equation \eqref{eq:abs_con} that the relative contrast $C$ follows a similar trend, thus explaining the decrease in the contrast with increasing readout duration, observed in Figure \ref{fig:ro_combined}(b).

Figure \ref{fig:noise_filter}(c) shows the amplitude response of the effective filter $\widetilde{H}(f)\widetilde{G}(f)$ for different readout durations, with the red and black curves corresponding to $T_{ro}=40 \mu s$ and $T_{ro}=100 \mu s$ respectively. It is observed that for a longer readout duration, the effective filter is narrower in frequency, and thus allows a smaller band of noise in the raw data to affect the processed data. This is consistent with the decrease in $\sigma_{odmr}$ - representative of the noise in the processed data -  with increasing readout duration, observed in Figure \ref{fig:ro_combined}(c). The noise PSD within the IF band which contributes most significantly to $\sigma_{odmr}$ is reasonably frequency independent for all values of $T_{ro}$ considered here, as seen in Figure \ref{fig:sig_noise}(b). Thus, assuming $\widetilde{S}_{\delta v}(f)$ to be white noise in equation \eqref{eq:sigma_odmr_psd}, we arrive at the proportionality relation of \eqref{eq:noise_tro}. This theoretical prediction captures the experimental trend in Figure \ref{fig:ro_combined}(c), as indicated by the black fit curve.

\begin{equation}
    \sigma^2_{odmr} \propto \frac{1}{T_{ro}}
    \label{eq:noise_tro} 
\end{equation}

The variation of the contrast to noise ratio with the readout time is shown in Figure \ref{fig:ro_combined}(d). The CNR is found to increase with increasing readout time, till around $260 \mu s$. Beyond this, any further increase in readout time does not change the CNR significantly. Notably, both the contrast and noise decrease when the  readout time is extended. The observed behavior of the CNR therefore arises from the interplay between these competing effects : an initial regime where the noise reduction dominates followed by a regime where the concurrent decrease in contrast offsets any further gain in noise suppression.

Increasing the readout duration increases the minimum waveform length $T_{wav}$ that must be captured on the oscilloscope. This, in turn, leads to a higher acquisition time, as shown in Appendix \ref{app:acq_oh}. Therefore, the readout duration should be chosen to be as short as possible, while ensuring an adequate CNR. A value of $T_{ro}=200 \mu s$ provides a suitable compromise, and is adopted for all other measurements.

\section{Impact of Sampling Rate and Noise Bandwidth}\label{sec:samp_rate}

\begin{figure*}
  \includegraphics[width=\linewidth]{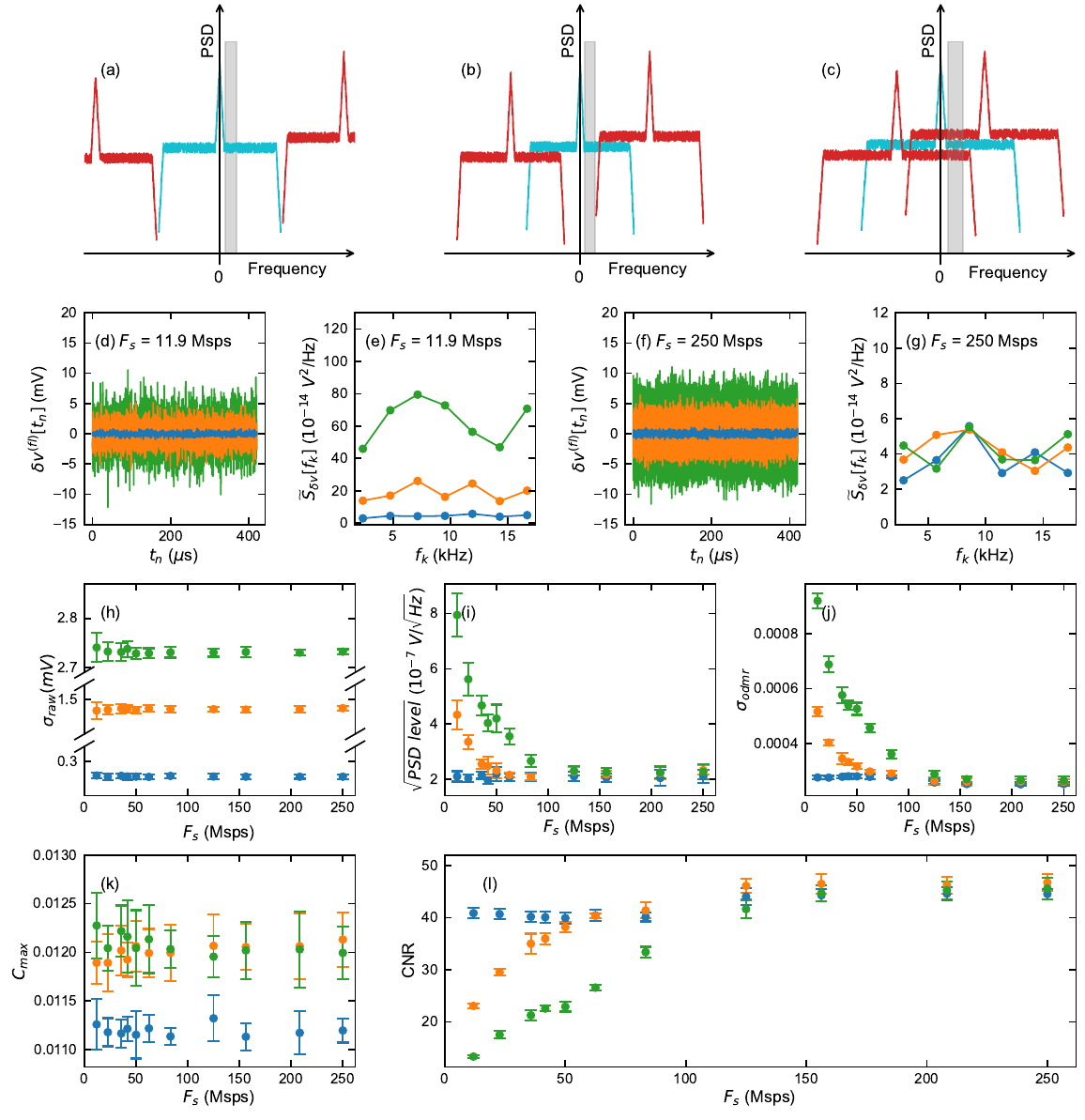}
\caption{\label{fig:samp_combined}\textbf{(a–c)} Schematic illustration of the analog noise PSD (cyan) and its aliased replicas (red). The gray region denotes the intermediate frequency (IF) band which is the primary contributor to the noise in the processed data.
 \textbf{(a)} Negligible overlap between spectral aliases.
\textbf{(b)} Substantial spectral overlap of the aliases, but outside the IF band.
\textbf{(c)} Substantial overlap of aliases within the IF band.
A consistent color scheme is used in figures \textbf{(d–l)}, where green, orange and blue represent configurations 1, 2 and 3 of the detection system respectively. \textbf{(d,f)} Time-domain intra-waveform noise for sampling rates $F_s=11.9$ Msps and $250 Msps$, respectively. 
\textbf{(e,g)} PSD within the IF band for the corresponding sampling rates.
\textbf{(h,i)} Variation of the standard deviation of the noise in raw data and the square root of PSD level in the IF band, respectively, with sampling rate for the three detection configurations.
\textbf{(j-l)} Variation of the noise in processed data, the ODMR contrast and the contrast to noise ratio, respectively, as a function of  sampling rate for the three detection configurations. These plots were obtained from data acquired with $T_{ro}=200 \mu s$, and $N_{avg}=32$. }
\end{figure*}

In this section, we introduce the concept of noise bandwidth of the detection system, and investigate its influence, together with the sampling rate, on the contrast to noise ratio (CNR) of the processed data. We further establish a condition relating the sampling rate and noise bandwidth that must be satisfied to achieve a high CNR.

This is particularly relevant to the oscilloscope-based data acquisition, as oscilloscopes typically operate at high default sampling rates, incurring substantial time and memory overheads, as discussed in Appendix \ref{app:acq_oh}. Thus, reducing the sampling rate can lead to faster acquisition, provided adequate CNR is maintained. We show that this can be achieved by modifying the detection system in a way which limits its noise bandwidth.

The aliasing effect, introduced in Section \ref{sec:sig_noise}, plays a key role in determining the dependence of CNR on the sampling rate and noise bandwidth. Figure \ref{fig:samp_combined}(a)-(c) schematically illustrates the relevant aliasing scenarios. The cyan curve represents the PSD of the analog noise, while the red curves represent two of its shifted replicas arising from aliasing during the sampling of the analog voltage by the ADC in the oscilloscope. According to equation \eqref{eq:aliasing},  the PSD of the digitized noise is given by the sum of the original analog noise PSD, and all its aliased components. The gray region denotes the intermediate frequency (IF) band, extending from $F_l$ to $F_u$, whose spectral components determine the noise in the processed ODMR data, as discussed in Section \ref{sec:noise_filt}.

Three distinct situations may arise depending on the sampling rate and the noise bandwidth. In the first case, the aliased spectra do not exhibit substantial overlap, as illustrated in Figure \ref{fig:samp_combined}(a). Under these conditions, noise aliasing is negligible. In the second case, substantial overlap between aliases occur, but entirely outside the IF band, as shown in Figure \ref{fig:samp_combined}(b). Although aliasing is present in the conventional sense, it doesn't affect the PSD within the IF region. In the third case, the aliased spectra overlap significantly within the IF band itself, as illustrated in Figure \ref{fig:samp_combined}(c). Only in the third situation does aliasing increase the PSD in the IF band, thereby increasing the noise in the processed data, as demonstrated later in this section.

We introduce a measure of the noise bandwidth which better captures the effect of aliasing, and accounts for the non-monotonic behavior that can sometimes occur in the PSD, as seen in Figure \ref{fig:sig_noise}(b). To ensure that aliasing has a minimal effect on the noise level within the IF band in the PSD of the digitized data, the PSD of the noise components aliased into this band should be at least one order of magnitude lower than the analog noise PSD within the IF band itself. To formalize this condition, we introduce the notion of the `relative $-10 dB $ noise bandwidth', denoted as $F_{rnbw}$, as the  frequency at which the analog noise PSD falls to a level $10 dB$ below its value within the IF band.

The conditions relating the sampling rate to $F_{rnbw}$ for the three situations illustrated in Figure \ref{fig:samp_combined}(a)-(c) are presented in Appendix \ref{app:alias_cond}. These results establish that the inequality in \eqref{strong_inequality} must be satisfied to achieve a good CNR. Specifically, this condition ensures that the aliasing of noise into the IF band, as shown in Figure \ref{fig:samp_combined}(c), is avoided, thereby preventing any increase in the noise of the processed ODMR data.
\begin{equation}
    F_s>F_{rnbw}+F_u
    \label{strong_inequality}
\end{equation}

For the chosen readout duration of $200 \mu s$, the resulting IF band is such that the condition $F_u<<F_s$ holds for all the sampling rates considered in this work. Consequently, the condition simplifies to the inequality \eqref{weak_inequality}, which requires the sampling rate to be higher than the noise bandwidth.

\begin{equation}
    F_s>F_{rnbw}
    \label{weak_inequality}
\end{equation}

It is noteworthy that these conditions are distinct from the Nyquist criterion, which is defined by the inequality \eqref{nyquist_inequality}.

\begin{equation}
    F_s>2F_{rnbw}
    \label{nyquist_inequality}
\end{equation}

Experimentally, the noise bandwidth is controlled by changing the configuration of the detection setup. Three configurations, labelled 1, 2 and 3, were defined in Section \ref{sec:methods} A. For configurations 1 and 2 where the signal is fed to the oscilloscope directly,  we find $F_{rnbw} \approx 102 MHz$ and $F_{rnbw} \approx 48 MHz$, respectively. In configuration 3 where the preamplifier acts as an analog low pass filter, the noise bandwidth is restricted to $F_{rnbw} \approx 1.68 MHz$. 

Figure \ref{fig:samp_combined}(d)-(i) present the experimental results for different combinations of sampling rate and noise bandwidth. A consistent color scheme is used throughout, with configurations 1, 2 and 3, corresponding to progressively lower noise bandwidths, shown in green, orange, and blue, respectively.

Figure \ref{fig:samp_combined}(h) shows the variation of $\sigma_{raw}$ with sampling rate for the different noise bandwidth configurations. For a given configuration, $\sigma_{raw}$ is approximately independent of the sampling rate. Configurations with larger noise bandwidth result in higher values of $\sigma_{raw}$, as the integral in equation \eqref{eq:sigma_raw_cont_psd} evaluates to a larger value. This trend is further illustrated in Figures \ref{fig:samp_combined}(d) and (f), which show the fluctuations in the raw data for $F_s=11.9 Msps$ and $F_s=250 Msps$, respectively.

The IF PSD level shows markedly different trends. This is evident from Figure \ref{fig:samp_combined}(i) which shows the variation of the square root of the IF PSD level with sampling rate for different configurations. At the highest sampling rates, the PSD level is low and almost identical across all the configurations. At lower sampling rates, configurations 1 and 2 show elevated PSD levels because the noise bandwidth exceeds the sampling rate, thereby violating the inequality \eqref{weak_inequality}. Since configuration 1 has a larger noise bandwidth, it experiences more severe aliasing, and therefore exhibits higher PSD levels than configuration 2. In contrast, configuration 3 shows lower PSD levels even at smaller sampling rates. This happens because configuration 3 has a substantially lower noise bandwidth, so condition in \eqref{weak_inequality} holds for all the sampling rates considered, ensuring that the effect of aliasing remains minimal. This is further illustrated in Figure \ref{fig:samp_combined}(e) and (g), where the noise power spectral density in the IF band is plotted for $F_s=11.9 Msps$ and $F_s=250 Msps$ respectively.

After examining how the sampling rate and noise bandwidth influence the temporal and spectral characteristics of the noise in the raw data, we now investigate their impact on the processed ODMR data. The maximum ODMR contrast is found to be independent of the sampling rate, as shown in Figure \ref{fig:samp_combined}(k). The contrast is slightly lower for configuration 3 due to a small DC offset introduced in the raw data by the preamplifier. The variation of $\sigma_{odmr}$ - the standard deviation of the noise in the processed data with sampling rate is shown in Figure \ref{fig:samp_combined}(j). It closely follows the trend of $\sqrt{\mathrm{PSD}\;\mathrm{level}}$, shown in Figure \ref{fig:samp_combined}(i). This is consistent with the discussion in Section \ref{sec:noise_filt}, which theoretically establishes that the noise in the IF band is the dominant contributor to $\sigma_{odmr}$. 

The variation of the contrast to noise ratio (CNR) with sampling rate is shown in Figure \ref{fig:samp_combined}(l). For configurations 1 and 2, the CNR decreases whenever the condition in inequality \eqref{weak_inequality} is violated. In contrast, configuration 3 satisfies this inequality over the entire range of sampling rates considered, thereby maintaining a consistently high CNR. These results demonstrate that the analog low pass filtering provided by the preamplifier enables high-CNR measurements even at low sampling rates. 

It is noteworthy that $\sigma_{odmr}$ is completely independent of $\sigma_{raw}$.
This demonstrates that the noise in the processed data cannot be inferred from the fluctuations in the raw data. Instead, the spectral components of the noise within the IF band must be examined. This further validates the DSP framework developed in this work.

\section{Discussion and Conclusions}

We have presented the complete workflow for the acquisition of the pulsed ODMR data using an oscilloscope, including both the acquisition of the raw waveform data, and the subsequent processing steps required to obtain the final ODMR contrast. The noise characteristics of the raw data have been investigated in detail. It has been established that the low frequency components of the noise lead to variations across successive waveforms, while the high frequency components result in  fluctuations within individual waveforms.

A key contribution of this work is the modeling of the data processing stage as a digital filter with transfer function $\widetilde{H}(f) \widetilde{G}(f) \widetilde{K}(f)$. The filter $\widetilde{H}(f) \widetilde{G}(f)$, which arises from the data processing for one iteration of the pulse sequence, exhibits a bandpass response. The filter $\widetilde{K}(f)$ which arises from averaging over multiple iterations of the pulse sequence, is found to effectively scale the noise in the processed data by a factor of $1/N_{avg}$. These results are broadly applicable to the absolute contrast measurements in pulsed ODMR, irrespective of the detection technique and material platform used.  This filter framework has enabled us to theoretically identify and quantify an intermediate frequency (IF) band whose noise spectral components make the dominant contribution to the noise in the processed ODMR data. Our experiments have confirmed that the noise in the processed data is determined primarily by the noise level in the IF band, and is independent of the total noise power present in the raw data.

Using these insights, the impact of various experimental parameters on the quality of the processed ODMR data, quantified by the contrast to noise ratio (CNR), has been investigated. During the data acquisition and processing,  two distinct forms of averaging take place: averaging within the signal and reference segments, characterized by the readout duration $T_{ro}$; and the averaging over waveforms acquired over successive iterations of the pulse sequence. These two operations affect the contrast to noise ratio (CNR) differently. While increasing the number of waveform averages continuously improves the CNR, increasing the $T_{ro}$ enhances CNR only up to a certain duration before it plateaus. The combined influence of sampling rate and analog noise bandwidth on the CNR has been analyzed. This analysis shows that placing an analog low pass filter before the oscilloscope improves acquisition efficiency by suppressing spectral aliasing.

A general prescription to be followed to ensure efficient pulsed ODMR acquisition has been presented in Appendix \ref{app:sop}. These guidelines have been successfully applied to the acquisition of Rabi Oscillations, Ramsey Interferometry, Spin echo and $T_1$- relaxation measurements, as demonstrated in the Supplementary Materials. We expect that the insights developed here will guide the design, and development of dedicated instruments for ODMR in future.

\begin{acknowledgments}
The authors sincerely acknowledge University Grant Commission,Department of Science and Technology(Grant No.DST/ICPS/Quest/2019/22) for funding this work.  A N acknowledges PMRF for research fellowship. H H acknowledges AICTE for fellowship. S M and S C acknowledge IISER Kolkata for funding. 

\end{acknowledgments}

\appendix

\section{Measurement of Standard Deviation and Allan deviation}\label{app:sd_ad}

\begin{figure}
 \includegraphics[]{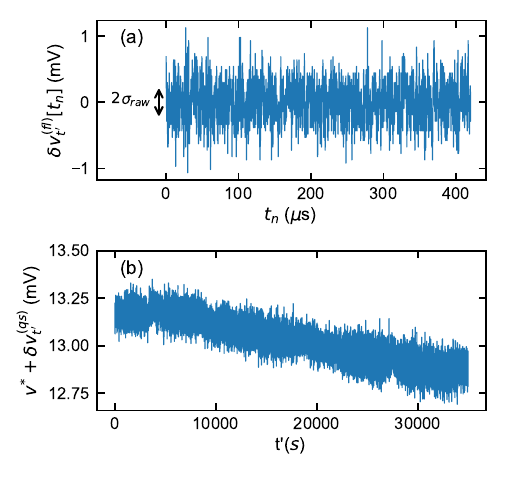}
 \centering
\caption{\label{fig:sigma_allan_raw}  
\textbf{(a)} The photovoltage fluctuations within a single waveform. $\sigma_{\mathrm{raw}}$ is the standard deviation of the waveform, which quantifies the noise in the raw data.  \textbf{(b)} Time series of mean voltage from successive waveforms triggered at 1 second intervals over an extended duration. This time series is used to calculate the Allan deviation.
}
\end{figure}

In this Appendix, we explain the method adopted for measuring the standard deviation and the Allan deviation.

The experiment is conducted in the absence of microwaves, while the sample is continuously illuminated by the laser. These conditions ensure that a constant red photoluminescence  would be obtained in the ideal case. Thus, the ideal noiseless photovoltage is independent of both $t'$ and $t_n$, i.e., $v^{*}_{t'}[t_n]=v^{*}$. So, equation \eqref{eq:sig_noise} and \eqref{eq:noise_comp} simplify to \eqref{eq:sig_noise_demo}. 

\begin{equation}
    v_{t'}[t_n] = v^{*}+\delta v^{(fl)}_{t'}[t_n]+\delta v^{(qs)}_{t'}
\label{eq:sig_noise_demo}
\end{equation}

The fluctuations within each waveform, $\delta v^{(fl)}_{t'}[t_n]$, are obtained by subtracting the waveform mean from each data point. A representative example is shown in Figure \ref{fig:sigma_allan_raw}(a), where the standard deviation of the fluctuations gives us $\sigma_{raw}$. The mean value of a given waveform then corresponds to the sum of the ideal photovoltage $v^{*}$, and the quasi-static noise contribution $\delta v_{t'}^{(qs)}$. The ideal noiseless voltage $v^{*}$ can be chosen arbitrarily, since the only requirement for $v^{*}$ is that it should remain constant throughout the experiment. This choice, however, doesn't change the value of the computed Allan deviation.


A large number of waveforms triggered at equal intervals of $1$ second are acquired, such that each waveform captures the same duration $T_{wav}$ as that used for pulsed ODMR data acquisition, shown in Figure \ref{fig:methods}(c). The variation in the corresponding $v^{*}+\delta v_{t'}^{(qs)}$ with the trigger time $t'$ is shown in Figure \ref{fig:sigma_allan_raw}. The overlapping Allan deviation of this time series is then estimated using standard methods\cite{riley2008handbook}. 

\section{Quantization Noise}\label{app:quant_noise}

The analog to digital converter (ADC) in any oscilloscope has a finite resolution determined by its number of bits. The oscilloscope uses a 10 bit ADC, corresponding to $2^{10}$ digital levels. Each sampled analog voltage is rounded to the closest digital level, resulting in a quantization error $\delta q[t_n]$. 

If the ADC resolution is $\Delta$, then the power spectral density (PSD) corresponding to the quantization noise is given by equation \eqref{eq:quant_psd}\cite{bennett1948spectra}.

\begin{equation}
   \widetilde{S}_{\delta q}(f)=\frac{\Delta^2}{12 F_s}
   \label{eq:quant_psd}
\end{equation}

In the present case, $\Delta = 19.5 \mu V$. For the sampling rates considered in this work, $\widetilde{S}_{\delta q}(f)$ is estimated to be of the order of $10^{-19}-10^{-18}$ $V^2/Hz$. However, within the intermediate frequency (IF) band, the measured PSD is consistently of the order of $10^{-14}$ $V^2/Hz$ or above, as shown in Section \ref{sec:sig_noise} and \ref{sec:samp_rate}. Quantization noise is therefore several orders of magnitude smaller than the measured noise, and can safely be ignored in our analysis.

\section{Estimation of Power Spectral Density}\label{app:psd_est}

In this Appendix, we describe the Average Periodogram method which is used for estimating the power spectral density from the oscilloscope waveform data.

For each waveform acquired at trigger time $t'$, the intra-waveform fluctuations $\delta v^{(fl)}_{t'}[t_n]$ is obtained using the method described in Appendix \ref{app:sd_ad}, with a representative example shown in Figure \ref{fig:sigma_allan_raw}(a). The discrete Fourier Transform (DFT) of $\delta v^{(fl)}_{t'}[t_n]$ is denoted as $\widetilde{V}^{(\delta)}_{t'}[f_k]$. The corresponding periodogram is calculated by normalizing the squared magnitude of the DFT according to equation \eqref{eq:periodogram} where $N$ is the number of samples in the waveform.

\begin{equation}
    \widetilde{S}_{\delta v_{t'}}[f_k]=\frac{\big|\widetilde{V}^{(\delta)}_{t'}[f_k]\big|^2}{N.F_s}
    \label{eq:periodogram}
\end{equation}

Repeating this procedure for waveforms acquired at different trigger times $t'$ yields a set of periodograms, $\{\widetilde{S}_{\delta v_{t'}}[f_k]\}$. The power spectral density is defined as the expectation value of the periodogram, and is estimated by averaging over this ensemble, as shown in equation \eqref{eq:psd_estimation}.

\begin{equation}
    \widetilde{S}_{\delta v}[f_k]=E\{\widetilde{S}_{\delta v_{t'}}[f_k] \} \approx \langle \widetilde{S}_{\delta v_{t'}}[f_k] \rangle
    \label{eq:psd_estimation}
\end{equation}

The accuracy of the PSD estimate improves with the number of periodograms used in the averaging. Throughout this work, the PSD is estimated by averaging over 32 periodograms.

For the calculation of the PSD's shown in Figure \ref{fig:sig_noise}(b), the waveforms are acquired at a very high sampling rate of $2.5 Gsps$. The power spectral density thus obtained provides a good estimate of the underlying analog PSD up to some spectral leakage.

\section{Derivation of Spectral Leakage Relation}\label{app:spec_leak}

Let $\delta v[t_n]$ be an infinitely long noise time series with autocorrelation $r_v[t_n]$, and let $\delta v^{(fin)}[t_n]$ be a finite segment of length $T_{wav}$ of this time series. Then, we can write 

\begin{equation}
    \delta v^{(fin)}[t_n]=w_v[t_n]\delta v[t_n]
\end{equation}

Here, $w_v[t_n]$ is a rectangular function extending from $0$ to $T_{wav}$, and can be written as 

\begin{equation}
    w_v[t_n]=u[t_n]u[T_{wav}-t_n]
\end{equation}

It has been shown in \cite{proakis2007digital} that the autocorrelation function gets modified as 

\begin{equation}
    r^{(fin)}_v[t_n]=w_r[t_n]r_v[t_n]
    \label{eq:auto__corr_fin}
\end{equation}

Here, $w_r[t_n]$ is a triangular function given by 

\begin{equation}
    w_r[t_n]=1-\frac{|t_n|}{T_{wav}}
\end{equation}

Taking DTFT on both sides of \eqref{eq:auto__corr_fin} and normalizing appropriately, we get 

\begin{equation}
    \widetilde{S}^{(fin)}_{\delta v}(f)=\widetilde{S}_{\delta v}(f)  \circledast \widetilde{W}(f)
\end{equation}

Here, 

\begin{equation}
    \widetilde{W}(f)=\frac{T_{wav}}{T_s} \big(\frac{sinc( f T_{wav})}{sinc( f T_s)}\big)^2
\end{equation}

Now, since $T_{wav}>>T_s$, within the range $-F_s/2$ to $+F_s/2$, we can approximate, 

\begin{equation}
    \widetilde{W}(f) \approx \frac{T_{wav}}{T_s} sinc^2( f T_{wav})
\end{equation}

Now, the DFT based PSD, $S_{\delta v}[f_k]$, is just the DTFT based PSD calculated at discrete frequency points.

\begin{equation}
    \widetilde{S}_{\delta v}[f_k]=\widetilde{S}^{(fin)}_{\delta v}(f=f_k)
\end{equation}

\section{Spectral contributions to Standard deviation and Allan Deviation}\label{app:spec_cont}

\begin{figure*}
    \centering
    \includegraphics[]{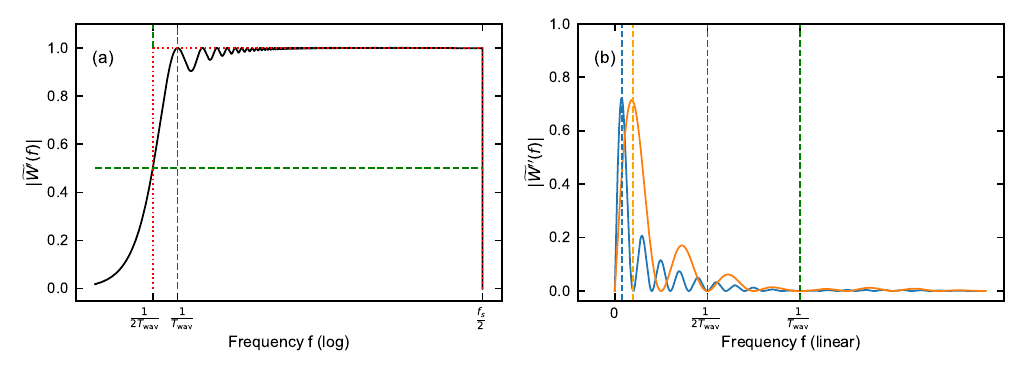}
    \caption{\textbf{(a)} Amplitude response of the spectral function $\widetilde{W}'(f)$. The green vertical lines are placed at $f=1/2T_{wav}$ and $f=1/T_{wav}$. The green horizontal line corresponds to $|\widetilde{W}'(f)|=0.5$. The spectral function is approximated by the red dotted rectangle.
    \textbf{(b)} Amplitude response of the spectral function of Allan deviation, $\widetilde{W}''(f)$. The blue and orange curves correspond to a longer  and a shorter  value of $\tau$ where the corresponding dotted vertical line shows the position of $f=1/\tau$.}
    \label{fig:app_spectral_cont}
\end{figure*}

The contributions of different spectral components of noise to the standard deviation and Allan deviation is discussed in Subsection C of Section \ref{sec:sig_noise}. In this Appendix, the underlying  framework is presented, thereby justifying the claims made in the main text.

The function $\widetilde{W}'(f)$ is given in equation \eqref{eq:w_prime}.

\begin{equation}
    \widetilde{W}'(f)=\sum_{n=1}^{M-1} \frac{T_s}{T_{wav}}\widetilde{W}\big(f-\frac{n}{T_{wav}}\big)
    \label{eq:w_prime}
\end{equation}

The amplitude response of $\widetilde{W}'(f)$ is shown in Figure \ref{fig:app_spectral_cont}(a). It oscillates close to $1$ within the range $1/T_{wav}<f<f_s/2 -1/T_{wav}$, while exhibiting a sharp attenuation outside this interval, i.e. for $f<1/T_{wav}$ and $f>F_s/2-1/T_{wav}$. There is a sharp reduction in $|\widetilde{W}'(f)|$ outside this range, i.e. below $f=1/T_{wav}$ and beyond $f=F_s/2 -1/T_{wav}$. In particular, it reaches the 3-dB point ($|\widetilde{W}'(f)|=0.5$) at $f=1/2T_{wav}$. This indicates that frequency components below $1/2T_{wav}$ contribute negligibly to the standard deviation. The same can be said for the spectral components above $f=F_s/2+1/2T_{wav}$. Therefore, $|\widetilde{W}'(f)|$ can be approximated by a rectangular function, as illustrated by the red dotted lines of Figure \ref{fig:app_spectral_cont}(a). Using this approximation, equation \eqref{eq:sigma_raw_cont_psd} can be simplified to equation \eqref{eq:sigma_raw_cont_psd_simp}.

\begin{equation}
    \sigma^2_{raw} \approx 2\int_{1/2T_{wav}}^{F_s/2-1/2T_{wav}}  \widetilde{S}_{\delta v}(f)  df
    \label{eq:sigma_raw_cont_psd_simp}
\end{equation}

The spectral function $\widetilde{W}''(f)$ for the Allan deviation is given by equation \eqref{eq:w_double_prime}.

\begin{equation}
    \widetilde{W}''(f)=\frac{sin^4(\pi f \tau)}{(\pi f \tau)^2} sinc^2(f T_{wav})
    \label{eq:w_double_prime}
\end{equation}

$|\widetilde{W}''(f)|$ for two values of $\tau$ are shown in Figure \ref{fig:app_spectral_cont}(b). The blue curve corresponds to the larger value of $\tau$, while the orange curve represents the smaller value. It is clear from the Figure that $|\widetilde{W}''(f)|$ attains its maximum value near $f \sim 1/\tau$, as shown by the dotted lines. It becomes negligibly small beyond $1/T_{wav}$. In practice, most of its spectral weight is concentrated below $1/2T_{wav}$, with only a small residual contribution in the range $1/2T_{wav}<f<1/T_{wav}$.

\section{Quantifying the Intermediate Frequency band}\label{app:if_band}

As discussed in Section \ref{sec:noise_filt}, the data processing operation is equivalent to a band pass filter with transfer function $\widetilde{H}(f) \widetilde{G}(f)$. Consequently, only noise components within an intermediate frequency (IF) band have substantial contribution to the noise in the processed ODMR data.  
While there are many possible ways to quantify this band, we use the definition presented in this Appendix throughout this work.

Let $F_0$ be the frequency at which the amplitude response of the filter attains its maximum value, $|\widetilde{H}(F_0) \widetilde{G}(F_0)|$. The lower edge of the IF band, $F_l$, is the smallest  frequency at which the amplitude response exceeds 10 percent of its highest value. The upper edge of the band, $F_u$, is the frequency beyond which the amplitude response is always lower than 10 percent of its highest value. Thus, IF band extends from $F_l$ to $F_u$, with the width of the band being $F_{if}=F_u-F_l$. This definition ensures that the noise spectral components outside this band are attenuated by at least a factor of 100 in power relative to the component at $F_0$.

Note that the band is different for different readout durations $T_{ro}$. The IF band defined in this way extends from $0.12 kHz$ to $18.2 kHz$.

\section{Variation of Contrast from Filter Function}\label{app:con_freq}

The absolute contrast $D$ is an element of the time series $d[t_n]$, as described in Section \ref{sec:noise_filt}. Let $\widetilde{V}(f)$ be the Discrete Time Fourier transform (DTFT) of the raw data $v[t_n]$. Then, the DTFT of $d[t_n]$ is $\widetilde{V}(f)\widetilde{H}(f)\widetilde{G}(f)$. Thus, $D$ is the inverse DTFT of $\widetilde{V}(f)\widetilde{H}(f)\widetilde{G}(f)$, evaluated at $t_n=T_{end}^{(sig)}$, as expressed in equation \eqref{eq:d_freq}.

\begin{align}
    D&=d[T_{end}^{(sig)}]\\
    &=\frac{1}{F_s}\int_{-F_s/2}^{F_s/2} \widetilde{V}(f) \widetilde{H}(f)\widetilde{G}(f) exp(i2 \pi f T_{end}^{(sig)}) df
    \label{eq:d_freq}
\end{align}

 On decomposing each term into its magnitude and phase part, and using the relation $T^{(sig)}_{start}=T_{end}^{(sig)}-T_{ro}$, we get the following simplified form.

\begin{equation}
    D=\frac{1}{F_s}\int_{-F_s/2}^{+F_s/2} |\widetilde{V}(f) \widetilde{H}(f)\widetilde{G}(f)|  exp(i \phi(f))  df
\end{equation}

where 

\begin{equation}
    \phi(f)=\angle \widetilde{V}(f)+\frac{\pi}{2}+ \pi f (2T_{start}^{(sig)} - T_{dark})
\end{equation}

Here, we have used the exact expressions for $\widetilde{H}(f)$ and $\widetilde{G}(f)$ from equations \eqref{eq:dirichlet} and \eqref{big_g} respectively.
The value of the integral is strongly influenced by the frequency-dependent phase factor $\phi(f)$. Each frequency component contributes as a complex phasor. Since $D$ is obtained by integrating these phasors over frequency, the individual contributions combine vectorially rather than algebraically. Consequently, constructive and destructive interference between contributions from different frequency components can occur. Hence, one cannot infer from the amplitude response of the filter ($|\widetilde{H}(f) \widetilde{G}(f)|$) alone, which frequency components contribute more or less to the contrast $D$.

\section{Noise PSD of the Averaged Waveform}\label{app:psd_wav_avg}

\begin{figure}
\centering
\includegraphics[]{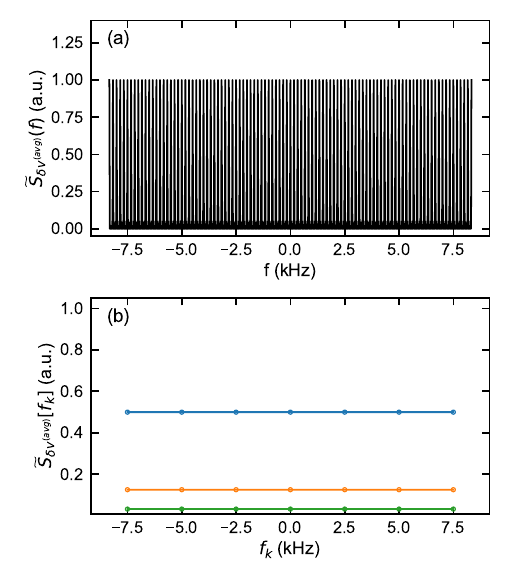}
\caption{\label{fig:combined_wav_avg_psd} \textbf{(a)} The PSD of the averaged waveform ($\widetilde{S}_{\delta v^{(avg)}}(f)$) for $N_{avg}=8$. when the unaveraged waveforms have white noise. \textbf{(b)} The discrete PSD for $N_{avg}=2,8,32$ shown in blue, orange and green, respectively.}
\end{figure}

The waveform averaging operation can be interpreted as a digital filter with impulse response $\widetilde{K}(f)$, defined by equation \eqref{eq:k_tilda} in Section \ref{sec:wav_avg}. Assuming a white noise input spectrum, $\widetilde{S}_{\delta v}(f)=1$, the output PSD reduces to $\widetilde{S}_{\delta v^{(avg)}}(f)=|\widetilde{K}(f)|^2$, as illustrated in Figure \ref{fig:combined_wav_avg_psd}(a) for $N_{avg}=8$.

 The finite waveform duration $T_{wav}$ leads to spectral leakage, which modifies the PSD according to equation \eqref{eq:spectral_leakage}. The resultant discrete PSD is calculated using the acquisition parameters for pulsed ODMR, $T_{wav}=420 \mu s$ and $T_{trig}=6 ms$. The calculated discrete PSDs for $N_{avg}=2,8,32$, are shown in blue, orange and green, respectively, in Figure \ref{fig:combined_wav_avg_psd}(b). Owing to the spectral leakage, the structure in the original PSD is smeared out, such that the waveform averaging produces an overall scaling of the PSD. This behavior is evident from Figure \ref{fig:combined_wav_avg_psd}(b), where the discrete PSD is well described by the approximate relation in equation  \eqref{eq:psd_wav_approx}. This also agrees with the experimentally obtained data, shown in Figure \ref{fig:wav_avg}(b), and its inset.

 \section{Aliasing Conditions}\label{app:alias_cond}

 In this Appendix, we mention the conditions which should be satisfied for each of the situations in Figure \ref{fig:samp_combined}(a)-(c).

 For Figure \ref{fig:samp_combined}(a) where the aliases do not overlap at all, the following condition must be satisfied. 

 \begin{equation}
     F_s>2 F_{rnbw}
 \end{equation}

 For Figure \ref{fig:samp_combined}(b) where the aliases overlap but not in the IF band, the following condition must be satisfied.

\begin{equation}
     2 F_{rnbw}>F_s>F_{rnbw}+F_u
 \end{equation}

For the aliases to overlap in the IF band, as shown in Figure \ref{fig:samp_combined}(c), the condition to be satisfied is

\begin{equation}
     F_s<F_{rnbw}+F_u
 \end{equation}

 \section{Calculation of Contrast to Noise Ratio}\label{app:calc_cnr}

This Appendix describes the procedure used to calculate the contrast to noise ratio from the pulsed ODMR data. Although the process is demonstrated for the relative contrast, it can be analogously applied to the absolute contrast.

\begin{figure}
\centering
\includegraphics[]{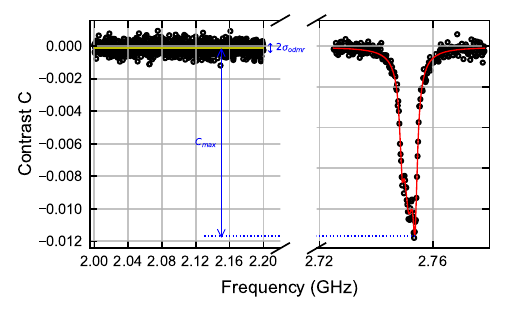}
\caption{\label{fig:demo_snr} Two data segments are shown. The segment to the left of the break captures a portion of the ODMR spectrum without any resonances. The standard deviation $\sigma_{odmr}$ of the fluctuations in this segment characterizes the noise in the processed data, while the yellow line represents its mean value. The segment shown to the right of the break contains the ODMR dips, with the red line showing the triple Lorentzian fit to the data. The difference between the lowest point of the fit and the mean noise level, as  indicated by $C_{max}$. For clarity, the x-axis of the right hand segment has been expanded, gridlines have been added to aid visualization.}
\end{figure}

The pulsed ODMR data are acquired over two microwave frequency ranges, as shown in Figure \ref{fig:demo_snr}. A $200$ MHz wide flat region of the ODMR spectrum containing only noise is recorded to the left of the break in Figure \ref{fig:demo_snr}. $\sigma_{odmr}$ is simply the standard deviation of the noise in this region. The second frequency range, shown to the right of the break, contains the ODMR dips of the NV centers. This portion of the spectrum is fitted to a sum of three Lorentzian curves, one for each hyperfine transition. The maximum contrast $C_{max}$ is defined as the difference between the baseline, taken as the mean of the noise-only region and the minimum of the fitted spectrum. The contrast to noise ratio is then defined as 

\begin{equation}
    CNR=\frac{C_{max}}{\sigma_{odmr}}
\end{equation}

\section{CNR for Absolute versus Relative Contrast}\label{app:comp_cnr}

The absolute and relative contrast are related as 

\begin{equation}
    C=\frac{D}{I_{ref}}
\end{equation}

Using standard error propagation, we get, 

\begin{equation}
    \frac{\sigma_{odmr}^{(C)}}{C} = \sqrt{\left(\frac{\sigma_{odmr}^{(D)}}{D}\right)^2+\left(\frac{\sigma_{ref}}{I_{ref}}\right)^2}
\end{equation}

Here, $\sigma_{ref}$ represents the standard deviation of the noise in $I_{ref}$. In all the cases considered in this work, $\sigma_{{ref}}<<I_{ref}$. Therefore, the relation simplifies to 

\begin{equation}
    \frac{C}{\sigma_{odmr}^{(C)}} \approx \frac{D}{\sigma_{odmr}^{(D)}}
\end{equation}

Thus, the contrast to noise ratio is essentially identical whether it is evaluated using the absolute contrast or the relative contrast. The validity of this approximation has been verified for all the CNR curves presented in this work.

\section{Standard Procedure for Optimal Pulsed ODMR Data Acquisition}\label{app:sop}

In most cases, the oscilloscope acquisition window can be set up in a way similar to Figure \ref{fig:methods}(c). The readout duration , $T_{ro}$, is selected following the procedure described in Section \ref{sec:readout} and should be kept identical across all experiments to ensure consistency of the measured data. The dark time $T_{dark}$ depends on the specific experiment, and may even vary during the course of a single experiment. Accordingly, the acquisition window must satisfy 

\begin{equation}
    T_{wav}>2T_{ro}+T^{(max)}_{dark}
\end{equation}

 where $T^{(max)}_{dark}$ is the maximum dark time required in the experiment. On appropriately adjusting the horizontal position on the oscilloscope, this condition ensures that sufficient data is available for both the signal and reference segments. $T_{wav}$ should not be chosen unnecessarily large, as it increases the waveform averaging time.  Configuration 3 of the detection part is used for the data acquisition. On the oscilloscope, the  sampling rate is set  higher than $F_{rnbw}$. The number of waveform averages is left as the freely adjustable parameter, enabling an appropriate trade-off between CNR and acquisition time.

\section{Acquisition Overheads} \label{app:acq_oh}

\begin{figure*}
 \includegraphics[]{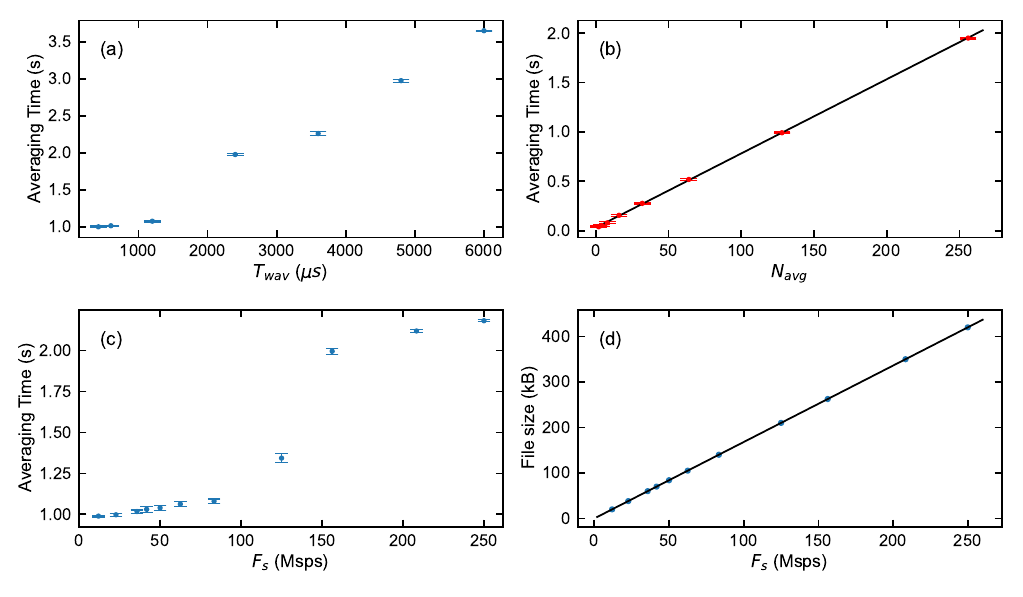}
\caption{\label{fig:resource_consumption}\textbf{(a-c)} The variation of the averaging time with window length, number of averages and sampling rate. In every plot the other two parameters are kept constant. \textbf{(d)} Variation of file size with sampling rate.} 
\end{figure*}

In this Appendix, we examine how the various experimental parameters like waveform length $T_{wav}$, number of averages $N_{avg}$, and sampling rate $F_s$ affect the  acquisition time. The primary bottleneck in the acquisition process was found to be the time required by the oscilloscope to perform waveform averaging. Therefore, we investigate the variation of the averaging time with these experimental parameters, as shown in Figure \ref{fig:resource_consumption}.

Figure \ref{fig:resource_consumption}(a) presents the variation of the averaging time with window length for $N_{avg}=128$ and $F_s=11.9 Msps$. The averaging time increases monotonically with the window length, and the same trend is observed for other parameter settings also. Figure \ref{fig:resource_consumption}(b) illustrates the variation of averaging time with number of averages for $F_s=11.9 Msps$ and $T_{wav}=420 \mu s$. The averaging time  increases approximately linearly with the number of averages as indicated by the black curve. Figure \ref{fig:resource_consumption}(c) shows the variation of the averaging time with the sampling rate for $N_{avg}=128$ and window length of $420 \mu s$. A monotonic increase in averaging time is again observed as the sampling rate is increased. While varying the window length and the number of averages do not alter the size of the data transmitted from the oscilloscope, increasing the sampling rate substantially increases data volume. As shown in Figure \ref{fig:resource_consumption}(d), the data size increases linearly with the sampling rate, thereby increasing the time and space overheads associated with subsequent processing and storage of the raw data.

\section*{Supplementary Materials}
\subsection{Power Spectral Density Definition}\label{app:psd_def}

The power spectral density (PSD) of the noise is defined via Wiener Khinchin theorem\cite{proakis2007digital}. For an analog noise process with autocorrelation function $r_{x}(t)$, the PSD is given by the Fourier transform of the autocorrelation function, as shown in equation \eqref{eq:psd_auto_corr}.

\begin{equation}
    S_{\delta x} (f)= \int_{-\infty}^{+\infty} r_{x}(t) e^{-i 2\pi f t} dt
    \label{eq:psd_auto_corr}
\end{equation}

For noise in the digital data, the autocorrelation is denoted as $r_{x}[t_n]$. The DTFT of $r_{ x}[t_n]$ then gives us a measure of the power spectral density\cite{proakis2007digital}. An additional factor of $1/F_s$ is introduced to ensure that $\widetilde{S}_{\delta x} (f)$ approaches the continuous-time PSD in the limit $F_s \rightarrow\infty$, and therefore has the same physical dimensions as $S_{\delta x} (f)$.

\begin{equation}
    \widetilde{S}_{\delta x} (f)= \frac{1}{F_s} \sum_{n=-\infty}^{+\infty} r_{x}[t_n] e^{-i 2\pi f t_n} 
    \label{eq:psd_auto_corr_dtft}
\end{equation}

In a similar way, $\widetilde{S}_{\delta x} [f_k]$ is obtained using the DFT of $r_{ x}[t_n]$ as shown in equation \eqref{eq:psd_auto_corr_dft}.

\begin{equation}
    \widetilde{S}_{\delta x} [f_k]= \frac{1}{F_s} \sum_{n=0}^{N-1} r_{x}[t_n] e^{-i 2\pi f_k t_n} 
    \label{eq:psd_auto_corr_dft}
\end{equation}

\subsection{Experimental Setup}

\begin{figure*}
 \includegraphics[]{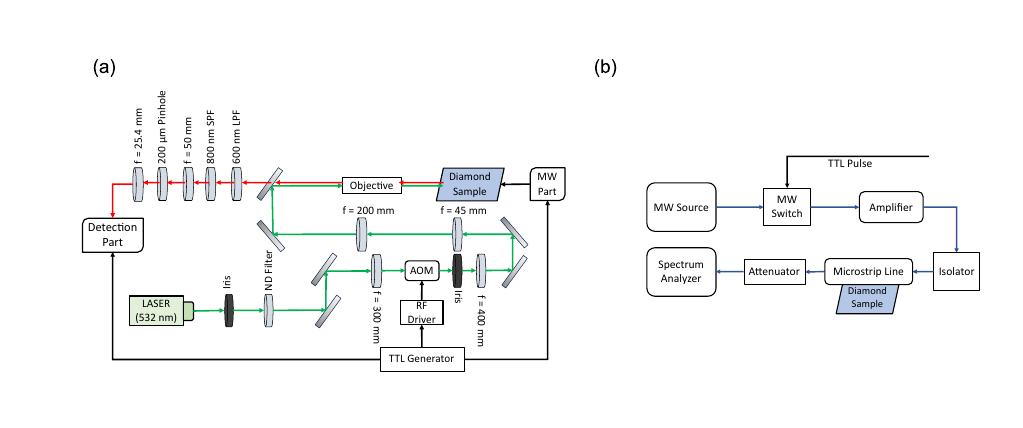}
\caption{\label{fig:SETUP}\textbf{(a)} Schematic of the complete experimental setup. The slanted rectangles represent mirrors. The grey circular shapes represent lenses and filters. The focal length of the lenses are written next to them. The cutoff wavelength of the wavelength filters are written next to them. SPF and LPF are abbreviations for short pass filter and long pass filter respectively. The black circular elements represent irises and pinholes. The green excitation laser is pulsed using an acousto-optic modulator (AOM). The laser is focused onto the diamond sample through the objective. The Red Photoluminescence coming from the sample is spectrally filtered using an 800~nm short-pass (SPF) and a 600~nm long-pass (LPF) filter, and focused into $200 \mu m$ pinhole and then directed to the detection system. The microwave part of the setup delivers the microwave to the sample.\textbf{(b)} Schematic of the microwave part of the experimental setup. The microwave coming from the source is pulsed using a MW switch, amplified, and delivered to a microstrip transmission line. The diamond sample is mounted on top of the microstrip line (MSL).} 
\end{figure*}

The experimental setup used for acquiring pulsed optically detected magnetic resonance (ODMR) is shown in Figure \ref{fig:SETUP}(a). The detection part and the microwave part of the circuit have been illustrated separately for the sake of clarity. The detection part has been described in Section III A and Figure 1(a) of the manuscript, while the microwave part is shown in Figure \ref{fig:SETUP}(b).
The diamond sample used in the experiment was obtained from Diamond Elements Pvt. Ltd. It was synthesized using chemical vapour deposition (CVD), subsequently irradiated with $2.5 MeV$ electrons, and annealed at $900^{\circ} \text{C}$. Although not explicitly shown in the Figures, a permanent neodymium magnet, placed close to the diamond sample, is used to generate the magnetic field.

The continuous laser light from a $532 nm$ source (Laser Quantum gem 532) is passed through an Acousto-Optic Modulator (Brimrose TEM-200-100-532) which is used for generating laser pulses. The AOM is placed at the focus of a telescopic arrangement of two plano-convex lenses, and is driven by an RF driver. The first order diffracted beam is chosen with an iris. By modulating the RF signal supplied to the AOM, the intensity of the first order diffracted beam is modulated to create laser pulses. The selected beam is subsequently expanded with a second telescopic arrangement to overfill the entrance aperture of a microscope objective lens (Newport M-40x, NA-0.65), which focuses the beam on to the sample. The laser power incident at the sample is $4 mW$. The light reflected from the sample is spectrally filtered to isolate the red photoluminescence emitted by the NV centers.
A 532 nm notch filter, as well as a $600 nm$ long pass filter (LPF) and a $800 nm$ short pass filter (SPF) are used. The red PL is passed through a $200 \mu m$ pinhole, thereby forming a confocal arrangement. It is subsequently focused into the detection part of the setup, shown in Figure 1(a) of the manuscript. 

The microwave part of the setup is shown in Figure \ref{fig:SETUP}(b). A continuous microwave is produced by a signal generator (Anapico APSIN4010), and is subsequently passed through the microwave switch (Mini-Circuits ZASWA-2-50-DR+) to form microwave pulses. The pulsed microwave signal is then amplified by the   Microwave amplifier (Mini-Circuits ZHL-16W-43-S+), and fed to a microstrip transmission line, which delivers the microwave magnetic field to the diamond sample kept on top of it. The microwave leaving the transmission line is appropriately attenuated and fed to the Spectrum Analyzer (Rohde and Schwarz FPC1500) for monitoring, and power measurements. 

The TTL generator (Pulseblaster ESR pro) supplies the TTL control pulses to the microwave switch, and the RF driver of the AOM, enabling the generation of microwave and laser pulses respectively. It also provides trigger signals for the oscilloscope, which ensures stable acquisition of the experimental data.

\subsection{Experimental Results}\label{app:exp_res}

\begin{figure*}
  \includegraphics[]{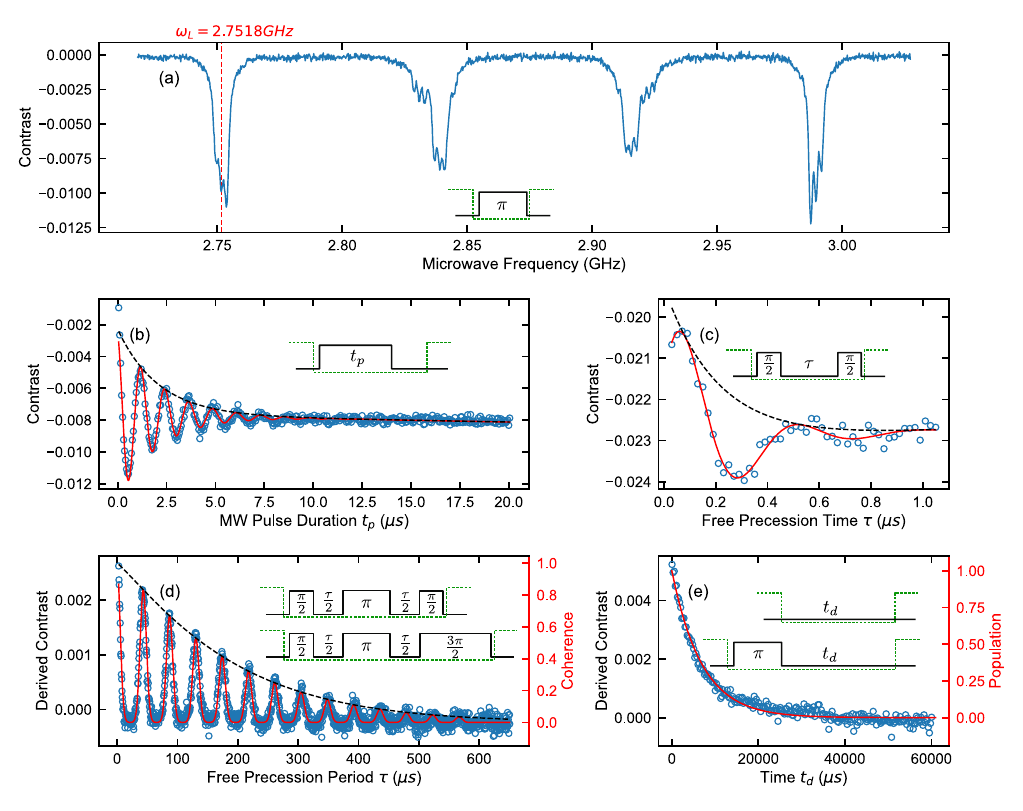}
\caption{\label{fig:encapsulated_results}\textbf{(a)} The full ODMR spectrum. A $\pi$-pulse is applied while the microwave frequency is swept.  The outermost resonance dips correspond to the NV centers approximately aligned with the applied magnetic field.\textbf{(b-e)} The blue circles represent the experimental data, while red lines show the corresponding fit. The black dotted lines in \textbf{(b-d)} indicate exponential decay envelopes.  \textbf{(b)} Rabi Oscillations Data. The duration of the microwave pulse is varied. \textbf{(c)} Ramsey Interferometry Data. The free precession time $\tau$ is varied. \textbf{(d)} Spin Echo data. The total free precession time $\tau$ is varied. \textbf{(e)} $T_1$ - decay data. The delay time $t_d$ is varied.} 
\end{figure*}

Figure \ref{fig:encapsulated_results} summarizes the experimental results obtained using the various pulsed ODMR based protocols. Figures \ref{fig:encapsulated_results} (a)-(e) present the full ODMR spectrum, Rabi Oscillations, Ramsey interferometry, spin echo and $T_1$-relaxation respectively. The corresponding microwave pulse sequence is shown alongside the plot corresponding to each measurement.

The relative contrast, expressed as a percentage, has been used to represent the full ODMR spectrum, Rabi Oscillations and  Ramsey Interferometry. On the other hand, the spin echo and the $T_1$-relaxation are represented by the derived contrast, which is defined as

\begin{equation}
    C_{der}=\frac{C_1-C_2}{2}
\end{equation}

Here, $C_1$ is the relative contrast obtained using the primary pulse sequence of each protocol (top sequence in Figure \ref{fig:encapsulated_results}(d) and (e)), and $C_2$ is the relative contrast obtained using the corresponding pulse sequence with an additional $\pi$ pulse (bottom sequence of Figure \ref{fig:encapsulated_results}(d) and (e)). Taking the difference between these two measurements suppresses contribution from the $T_1$ relaxation of the NV spins oriented along the other three crystallographic axes, as well as other dark time artefacts\cite{babashah2023optically}.

Figure \ref{fig:encapsulated_results}(a) shows the full pulsed ODMR spectrum. The outermost dips correspond to the NV centers whose symmetry axes are approximately aligned with the applied magnetic field. The central hyperfine transition of the leftmost dip at $\omega_L=2.7518 GHz$ is selected for all subsequent experiments, as indicated by the vertical red dotted line. 
The Rabi Oscillation data in Figure \ref{fig:encapsulated_results}(b) are fitted to extract a Rabi frequency of $\Omega_R=814.14 \pm 13.13 $ $kHz$, and a decay time $T_{2}^{(rabi)}$ of $2.07 \pm 0.06$ $\mu s$. Figure \ref{fig:encapsulated_results}(c) shows the Ramsey interference signal, which is well described by an exponentially decaying sinusoid of frequency $2.20 \pm 0.06$ $MHz$ and an inhomogeneous dephasing time of $T_{2}^{*}=202.8\pm 16.6 ns$. The spin echo data of Figure \ref{fig:encapsulated_results}(d) exhibit  periodic collapse and revivals of the coherence at the Larmor frequency of the carbon-13 nuclear spins. The revival amplitudes decay with an exponential envelope, yielding a coherence time of $T_2=195.13 \pm 3.34$ $\mu s$ . Finally, fitting the longitudinal relaxation data of Figure \ref{fig:encapsulated_results}(e) gives a depolarization time of $T_1=6.89 \pm 0.15$ $ms$.

\subsection{Data Visualization}
This video\cite{nandi2026rabi} illustrates the real-time visualization and acquisition of Rabi oscillations from an ensemble of NV centers in diamond using an oscilloscope. On the oscilloscope display, the blue waveform represents the red photoluminescence (PL) signal, averaged over 128 acquisitions, as indicated in the upper-right corner of the display. The yellow waveform corresponds to the microwave pulse whose duration is incrementally increased. For each acquired waveform, the corresponding contrast is calculated and added to the plot shown at the bottom. Initially, at very low microwave pulse durations, the signal segment of the PL waveform (right) is at a level similar to that as the reference segment (left). As the microwave pulse duration is gradually increased till the $\pi$ pulse length, the signal level gradually decreases relative to the reference segment. In this way, the oscilloscope provides real-time visualization of the data, enabling rapid identification and debugging of anomalies during the experiment.


\bibliography{apssamp}

@PREAMBLE{
 "\providecommand{\noopsort}[1]{}" 
 # "\providecommand{\singleletter}[1]{#1}%" 
}

@article{childress2014atom,
  title={Atom-like crystal defects: From quantum computers to biological sensors},
  author={Childress, Lilian and Walsworth, Ronald and Lukin, Mikhail},
  journal={Physics Today},
  volume={67},
  number={10},
  pages={38--43},
  year={2014},
  publisher={AIP Publishing}
}

@article{Balasubramanian2019,
  author  = {Balasubramanian, Priya and Osterkamp, Christoph and Chen, Y. and Chen, X. and Teraji, T. and Wu, E. and Naydenov, B. and Jelezko, F.},
  title   = {dc Magnetometry with Engineered Nitrogen-Vacancy Spin Ensembles in Diamond},
  journal = {Nano Letters},
  volume  = {19},
  number  = {9},
  pages   = {6681--6686},
  year    = {2019},
  doi     = {10.1021/acs.nanolett.9b02993}
}

@article{Michl2019,
  author  = {Michl, J. and Steiner, J. and Denisenko, A. and B{\"u}lau, A. and Zimmermann, A. and Nakamura, K. and Sumiya, H. and Onoda, S. and Neumann, P. and Isoya, J. and Wrachtrup, J.},
  title   = {Robust and Accurate Electric Field Sensing with Solid State Spin Ensembles},
  journal = {Nano Letters},
  volume  = {19},
  number  = {8},
  pages   = {4904--4910},
  year    = {2019},
  doi     = {10.1021/acs.nanolett.9b00900}
}

@article{Choe2018,
  author  = {Choe, J. and others},
  title   = {Precise temperature sensing with nanoscale thermal sensors based on diamond NV centers},
  journal = {Current Applied Physics},
  volume  = {18},
  number  = {9},
  pages   = {1066--1070},
  year    = {2018},
  doi     = {10.1016/j.cap.2018.06.002}
}

@article{Allert2022,
  author  = {Allert, Robin D. and Briegel, Karl D. and Bucher, Dominik B.},
  title   = {Advances in nano- and microscale NMR spectroscopy using diamond quantum sensors},
  journal = {Chemical Communications},
  volume  = {58},
  number  = {59},
  pages   = {8165--8181},
  year    = {2022},
  doi     = {10.1039/D2CC01546C}
}

@article{seo2017designing,
  title={Designing defect-based qubit candidates in wide-gap binary semiconductors for solid-state quantum technologies},
  author={Seo, Hosung and Ma, He and Govoni, Marco and Galli, Giulia},
  journal={Physical Review Materials},
  volume={1},
  number={7},
  pages={075002},
  year={2017},
  publisher={APS}
}

@article{awschalom2018quantum,
  title={Quantum technologies with optically interfaced solid-state spins},
  author={Awschalom, David D and Hanson, Ronald and Wrachtrup, J{\"o}rg and Zhou, Brian B},
  journal={Nature Photonics},
  volume={12},
  number={9},
  pages={516--527},
  year={2018},
  publisher={Nature Publishing Group UK London}
}

@article{ruf2021quantum,
  title={Quantum networks based on color centers in diamond},
  author={Ruf, Maximilian and Wan, Noel H and Choi, Hyeongrak and Englund, Dirk and Hanson, Ronald},
  journal={Journal of Applied Physics},
  volume={130},
  number={7},
  year={2021},
  publisher={AIP Publishing}
}

@article{aharonovich2022quantum,
  title={Quantum emitters in hexagonal boron nitride},
  author={Aharonovich, Igor and Tetienne, Jean-Philippe and Toth, Milos},
  journal={Nano letters},
  volume={22},
  number={23},
  pages={9227--9235},
  year={2022},
  publisher={ACS Publications}
}

@article{castelletto2020silicon,
  title={Silicon carbide color centers for quantum applications},
  author={Castelletto, Stefania and Boretti, Alberto},
  journal={Journal of Physics: Photonics},
  volume={2},
  number={2},
  pages={022001},
  year={2020},
  publisher={IOP Publishing}
}

@article{khoury2022bright,
  title={A bright future for silicon in quantum technologies},
  author={Khoury, Mario and Abbarchi, Marco},
  journal={Journal of Applied Physics},
  volume={131},
  number={20},
  year={2022},
  publisher={AIP Publishing}
}

@article{sewani2020coherent,
  title={Coherent control of NV- centers in diamond in a quantum teaching lab},
  author={Sewani, Vikas K and Vallabhapurapu, Hyma H and Yang, Yang and Firgau, Hannes R and Adambukulam, Chris and Johnson, Brett C and Pla, Jarryd J and Laucht, Arne},
  journal={American Journal of Physics},
  volume={88},
  number={12},
  pages={1156--1169},
  year={2020},
  publisher={AIP Publishing}
}

@article{nizovtsev2005quantum,
  title={A quantum computer based on NV centers in diamond: optically detected nutations of single electron and nuclear spins},
  author={Nizovtsev, AP and Kilin, S Ya and Jelezko, F and Gaebal, T and Popa, Iulian and Gruber, A and Wrachtrup, Jorg},
  journal={Optics and spectroscopy},
  volume={99},
  pages={233--244},
  year={2005},
  publisher={Springer}
}

@article{humphreys2018deterministic,
  title={Deterministic delivery of remote entanglement on a quantum network},
  author={Humphreys, Peter C and Kalb, Norbert and Morits, Jaco PJ and Schouten, Raymond N and Vermeulen, Raymond FL and Twitchen, Daniel J and Markham, Matthew and Hanson, Ronald},
  journal={Nature},
  volume={558},
  number={7709},
  pages={268--273},
  year={2018},
  publisher={Nature Publishing Group UK London}
}

@article{acosta2009diamonds,
  title={Diamonds with a high density of nitrogen-vacancy centers for magnetometry applications},
  author={Acosta, Victor M and Bauch, Erik and Ledbetter, Micah P and Santori, Charles and Fu, K-MC and Barclay, Paul E and Beausoleil, Raymond G and Linget, H{\'e}lo{\"\i}se and Roch, Jean Francois and Treussart, Francois and others},
  journal={Physical Review B—Condensed Matter and Materials Physics},
  volume={80},
  number={11},
  pages={115202},
  year={2009},
  publisher={APS}
}

@article{schirhagl2014nitrogen,
  title={Nitrogen-vacancy centers in diamond: nanoscale sensors for physics and biology},
  author={Schirhagl, Romana and Chang, Kevin and Loretz, Michael and Degen, Christian L},
  journal={Annual review of physical chemistry},
  volume={65},
  number={1},
  pages={83--105},
  year={2014},
  publisher={Annual Reviews}
}

@article{ho2022diamond,
  title={Diamond quantum sensors: from physics to applications on condensed matter research},
  author={Ho, Kin On and Shen, Yang and Pang, Yiu Yung and Leung, Wai Kuen and Zhao, Nan and Yang, Sen},
  journal={Functional Diamond},
  volume={1},
  number={1},
  pages={160--173},
  year={2022},
  publisher={Taylor \& Francis}
}

@article{basu2025diamond,
  title={Diamond color center based quantum metrology in industries for energy applications},
  author={Basu, Tanmoy and Patra, Anupam and Murali, Midhun and Saini, Mahesh and Banerjee, Amit and Som, Tapobrata},
  journal={ACS omega},
  volume={10},
  number={3},
  pages={2372--2392},
  year={2025},
  publisher={ACS Publications}
}

@article{wu2025fiber,
  title={Fiber-integrated NV Magnetometer with Microcontroller-based Software Lock-in Technique},
  author={Wu, Qilong and Shen, Xuan-Ming and Zhang, Yuan and Shan, Ying-Geng and Yu, Hui-Hui and Zhang, Jing-Hao and Chen, Jiahui and Wang, Yan and Yang, Xun and Tian, Yong-Zhi and others},
  journal={arXiv preprint arXiv:2510.01996},
  year={2025}
}

@article{pogorzelski2024compact,
  title={Compact and fully integrated LED quantum sensor based on NV centers in diamond},
  author={Pogorzelski, Jens and Horsthemke, Ludwig and Homrighausen, Jonas and Stiegek{\"o}tter, Dennis and Gregor, Markus and Gl{\"o}sek{\"o}tter, Peter},
  journal={Sensors},
  volume={24},
  number={3},
  pages={743},
  year={2024},
  publisher={MDPI}
}

@article{hiroshige2023compact,
  title={Compact and portable quantum sensor module using diamond NV centers [J]},
  author={Hiroshige, D and Tsukasa, H and Hiroya, S and others},
  journal={Applied Physics Express},
  volume={16},
  number={6},
  pages={06},
  year={2023}
}

@incollection{bar2019nv,
  title={NV color centers in diamond as a platform for quantum thermodynamics},
  author={Bar-Gill, Nir},
  booktitle={Thermodynamics in the Quantum Regime: Fundamental Aspects and New Directions},
  pages={983--998},
  year={2019},
  publisher={Springer}
}

@article{klatzow2019experimental,
  title={Experimental demonstration of quantum effects in the operation of microscopic heat engines},
  author={Klatzow, James and Becker, Jonas N and Ledingham, Patrick M and Weinzetl, Christian and Kaczmarek, Krzysztof T and Saunders, Dylan J and Nunn, Joshua and Walmsley, Ian A and Uzdin, Raam and Poem, Eilon},
  journal={Physical Review Letters},
  volume={122},
  number={11},
  pages={110601},
  year={2019},
  publisher={APS}
}

@article{choi2017observation,
  title={Observation of discrete time-crystalline order in a disordered dipolar many-body system},
  author={Choi, Soonwon and Choi, Joonhee and Landig, Renate and Kucsko, Georg and Zhou, Hengyun and Isoya, Junichi and Jelezko, Fedor and Onoda, Shinobu and Sumiya, Hitoshi and Khemani, Vedika and others},
  journal={Nature},
  volume={543},
  number={7644},
  pages={221--225},
  year={2017},
  publisher={Nature Publishing Group UK London}
}

@article{he2023quasi,
  title={Quasi-floquet prethermalization in a disordered dipolar spin ensemble in diamond},
  author={He, Guanghui and Ye, Bingtian and Gong, Ruotian and Liu, Zhongyuan and Murch, Kater W and Yao, Norman Y and Zu, Chong},
  journal={Physical Review Letters},
  volume={131},
  number={13},
  pages={130401},
  year={2023},
  publisher={APS}
}

@article{wu2025spin,
  title={Spin squeezing in an ensemble of nitrogen--vacancy centres in diamond},
  author={Wu, Weijie and Davis, Emily J and Hughes, Lillian B and Ye, Bingtian and Wang, Zilin and Kufel, Dominik and Ono, Tasuku and Meynell, Simon A and Block, Maxwell and Liu, Che and others},
  journal={Nature},
  volume={646},
  number={8083},
  pages={74--80},
  year={2025},
  publisher={Nature Publishing Group UK London}
}

@article{qiu2015quantum,
  title={Quantum Zeno and Zeno-like effects in nitrogen vacancy centers},
  author={Qiu, Jing and Wang, Yang-Yang and Yin, Zhang-Qi and Zhang, Mei and Ai, Qing and Deng, Fu-Guo},
  journal={Scientific Reports},
  volume={5},
  number={1},
  pages={17615},
  year={2015},
  publisher={Nature Publishing Group UK London}
}

@article{stanwix2010coherence,
  title={Coherence of nitrogen-vacancy electronic spin ensembles in diamond},
  author={Stanwix, Paul L and Pham, Linh My and Maze, Jeronimo R and Le Sage, David and Yeung, Tsun Kwan and Cappellaro, Paola and Hemmer, Philip R and Yacoby, Amir and Lukin, Mikhail D and Walsworth, Ronald L},
  journal={Physical Review B—Condensed Matter and Materials Physics},
  volume={82},
  number={20},
  pages={201201},
  year={2010},
  publisher={APS}
}

@article{suter2017single,
  title={Single-spin magnetic resonance in the nitrogen-vacancy center of diamond},
  author={Suter, Dieter and Jelezko, Fedor},
  journal={Progress in nuclear magnetic resonance spectroscopy},
  volume={98},
  pages={50--62},
  year={2017},
  publisher={Elsevier}
}

@article{jarmola2012temperature,
  title={Temperature-and magnetic-field-dependent longitudinal spin relaxation in nitrogen-vacancy ensembles in diamond},
  author={Jarmola, A and Acosta, VM and Jensen, K and Chemerisov, S and Budker, D},
  journal={Physical review letters},
  volume={108},
  number={19},
  pages={197601},
  year={2012},
  publisher={APS}
}

@article{mrozek2015longitudinal,
  title={Longitudinal spin relaxation in nitrogen-vacancy ensembles in diamond},
  author={Mr{\'o}zek, Mariusz and Rudnicki, Daniel and Kehayias, Pauli and Jarmola, Andrey and Budker, Dmitry and Gawlik, Wojciech},
  journal={EPJ Quantum Technology},
  volume={2},
  pages={1--11},
  year={2015},
  publisher={Springer}
}

@article{gorrini2021long,
  title={Long-lived ensembles of shallow NV--centers in flat and nanostructured diamonds by photoconversion},
  author={Gorrini, Federico and Dorigoni, Carla and Olivares-Postigo, Domingo and Giri, Rakshyakar and Apr\`a, Pietro and Picollo, Federico and Bifone, Angelo},
  journal={ACS Applied Materials \& Interfaces},
  volume={13},
  number={36},
  pages={43221--43232},
  year={2021},
  publisher={ACS Publications}
}

@article{masuyama2018extending,
  title={Extending coherence time of macro-scale diamond magnetometer by dynamical decoupling with coplanar waveguide resonator},
  author={Masuyama, Y and Mizuno, K and Ozawa, H and Ishiwata, H and Hatano, Y and Ohshima, T and Iwasaki, T and Hatano, M},
  journal={Review of Scientific Instruments},
  volume={89},
  number={12},
  year={2018},
  publisher={AIP Publishing}
}

@article{zhang2019modified,
  title={A modified spin pulsed readout method for NV center ensembles reducing optical noise},
  author={Zhang, Jixing and Yuan, Heng and Zhang, Ning and Xu, Lixia and Bian, Guodong and Fan, Pengcheng and Li, Mingxin and Zhan, Zhi and Yu, Kan},
  journal={IEEE transactions on instrumentation and measurement},
  volume={69},
  number={7},
  pages={4370--4378},
  year={2019},
  publisher={IEEE}
}

@book{graeme1995photodiode,
  title={Photodiode amplifiers: op amp solutions},
  author={Graeme, Jerald},
  year={1995},
  publisher={McGraw-Hill, Inc.}
}

@article{Kirchner2005,
  author = {Kirchner, James W.},
  title = {Aliasing in 1/f Noise Spectra: Origins, Consequences, and Remedies},
  journal = {Phys. Rev. E},
  volume = {71},
  issue = {6},
  pages = {066110},
  year = {2005},
  month = {Jun},
  publisher = {American Physical Society},
  doi = {10.1103/PhysRevE.71.066110}
}

@techreport{IEEE1139,
  author      = {{IEEE}},
  title       = {IEEE Standard Definitions of Physical Quantities for Fundamental Frequency and Time Metrology---Random Instabilities},
  institution = {IEEE},
  number      = {IEEE Std 1139-2008},
  year        = {2009},
  month       = mar,
}

@book{proakis2007digital,
  title={Digital signal processing: principles, algorithms, and applications, 4/E},
  author={Proakis, John G},
  year={2007},
  publisher={Pearson Education India}
}

@article{babashah2023optically,
  title={Optically detected magnetic resonance with an open source platform},
  author={Babashah, Hossein and Shirzad, Hoda and Losero, Elena and Goblot, Valentin and Galland, Christophe and Chipaux, Mayeul},
  journal={SciPost Physics Core},
  volume={6},
  number={4},
  pages={065},
  year={2023}
}

@article{aslam2013photo,
  title={Photo-induced ionization dynamics of the nitrogen vacancy defect in diamond investigated by single-shot charge state detection},
  author={Aslam, Nabeel and Waldherr, Gerhald and Neumann, Philipp and Jelezko, Fedor and Wrachtrup, Joerg},
  journal={New Journal of Physics},
  volume={15},
  number={1},
  pages={013064},
  year={2013},
  publisher={IOP Publishing}
}

@article{riley2008handbook,
  title={Handbook of Frequency},
  author={Riley, W},
  journal={NIST: Gaithersburg, MD, USA},
  year={2008}
}

@article{babunts2023magnetic,
  title={Magnetic Resonance Express Analysis and Control of NV- Diamond Wafers for Quantum Technologies},
  author={Babunts, Roman A and Gurin, Aleksandr S and Uspenskaya, Yulia A and Likhachev, Kirill V and Yakovleva, Valentina V and Batueva, Anastasia V and Skomorokhov, Artyom M and Veyshtort, Igor P and Uchaev, Maxim V and Eseev, Marat K and others},
  journal={Applied Magnetic Resonance},
  pages={1--12},
  year={2023},
  publisher={Springer}}

@article{bennett1948spectra,
  title={Spectra of quantized signals},
  author={Bennett, William Ralph},
  journal={The Bell System Technical Journal},
  volume={27},
  number={3},
  pages={446--472},
  year={1948},
  publisher={Nokia Bell Labs}
}

@misc{nandi2026rabi,
  author       = {Nandi, Anuvab and
                  Chakraborty, Sayan and
                  Jain, Yashvardhan and
                  Sharma, Kanav and
                  Mondal, Abir and
                  Chakraborti, Samiran and
                  Himadri, Himadri and
                  Mitra, Chiranjib},
  title        = {Real-Time Observation of Rabi Oscillations
                  using an Oscilloscope},
  year         = {2026},
  publisher    = {Zenodo},
  doi          = {https://doi.org/10.5281/zenodo.22869021},
  url          = {https://doi.org/10.5281/zenodo.22869021},
  howpublished = {Video}
}
\clearpage

\end{document}